\documentclass[aps,prd,10pt,notitlepage,nofootinbib]{revtex4-1}
\usepackage[utf8]{inputenc}
\usepackage{amsmath,amssymb,amsfonts}
\usepackage{newtxtext,newtxmath}

\usepackage{bbold}
\usepackage{bm}
\usepackage{graphicx}
\usepackage[usenames,dvipsnames]{xcolor}
\usepackage{color}
\usepackage[colorlinks=true,linkcolor=blue,urlcolor=blue,citecolor=blue]{hyperref}
\usepackage{slashed}
\usepackage[english]{babel}
\usepackage{dcolumn}
\usepackage{pifont}
\usepackage{dsfont,mathrsfs}
\usepackage{cancel}
\usepackage{bigints}
\usepackage{accents}
\usepackage{soul}
\usepackage{multirow}

\usepackage{makecell}
\usepackage{simpler-wick}
\newcommand{\nn}{\nonumber}
\newcommand{\ket}[1]{\left|#1\right\rangle}

\newcommand{\expectationvalue}[3]{\langle#1|#2|#3\rangle}
\newcommand{\ExpectationValue}[3]{\left\langle#1\left|#2\right|#3\right\rangle}
\newcommand{\ensembleaverage}[1]{\left\langle#1\right\rangle}

\newcommand{\MB}[1]{\left|#1\right|}

\newcommand{\FB}[1]{\left(#1\right)}
\newcommand{\fb}[1]{(#1)}
\newcommand{\SB}[1]{\left\{#1\right\}}
\newcommand{\TB}[1]{\left[#1\right]}
\newcommand{\AB}[1]{\left<#1\right>}

\newcommand{\mcT}{\mathcal{T}}

\newcommand{\mcI}{\mathcal{I}}
\newcommand{\mcF}{\mathcal{F}}
\newcommand{\mcSh}{\hat{\mathcal{S}}}
\newcommand{\mcZ}{\mathcal{Z}}

\newcommand{\mtW}{\mathtt{W}}
\newcommand{\mtL}{\mathtt{L}}
\newcommand{\msN}{\mathscr{N}}

\newcommand{\scrL}{\mathscr{L}}

\newcommand{\scrD}{\mathscr{D}}

\newcommand{\munu}{{\mu\nu}}

\newcommand{\IM}{\text{Im}}

\newcommand{\Tr}{\text{Tr}}

\newcommand{\psibar}{\overline{\psi}}

\newcommand{\del}{\partial}

\newcommand{\fsl}{\slashed}
\newcommand{\kfourint}[1]{\int \dfrac{d^4#1}{(2\pi)^4}}

\newcommand{\kthreeint}[1]{\int \dfrac{d^3\vec{#1}}{(2\pi)^3}}

\newcommand{\mcP}{{\mathcal{P}}}

\newcommand{\pvec}{\MB{\vec{p}}}
\newcommand{\qvec}{\MB{\vec{q}}}
\newcommand{\kvec}{|\vec{k}|}

\newcommand{\wpr}{\omega_{p}^r}
\newcommand{\wps}{\omega_{p}^s}
\newcommand{\wkr}{\omega_{k}^r}

\newcommand{\td}{\text{d}}

\begin{document}
	%\begin{linenumbers}
	\title{Dilepton production from chirally asymmetric matter}

	\author{Nilanjan Chaudhuri$^{a,d}$}
	\email{sovon.nilanjan@gmail.com}
	\email{n.chaudhri@vecc.gov.in}
	\author{Snigdha Ghosh$^{b}$}
	\email{snigdha.physics@gmail.com}
	\thanks{Corresponding Author}
	\author{Sourav Sarkar$^{a,d}$}
	\email{sourav@vecc.gov.in}
	\author{ Pradip Roy$^{c,d}$}
	\email{pradipk.roy@saha.ac.in}
	
	\affiliation{$^a$Variable Energy Cyclotron Centre, 1/AF Bidhannagar, Kolkata 700 064, India}
	\affiliation{$^b$Government General Degree College Kharagpur-II, Paschim Medinipur - 721149, West Bengal, India}
	\affiliation{$^c$Saha Institute of Nuclear Physics, 1/AF Bidhannagar, Kolkata - 700064, India}
	\affiliation{$^d$Homi Bhabha National Institute, Training School Complex, Anushaktinagar, Mumbai - 400094, India}

	%+++++++++++++++++++++++++++++++++++++++++++++++++++++++++++++++++++++++++++++++++++++++++++++++++++++++++++++++++++++
\begin{abstract}
We evaluate the dilepton production rate (DPR) from hot and dense chirally asymmetric quark matter. The presence of a finite chiral chemical potential (CCP) in the electromagnetic spectral function results in the appearance of new cut structures signifying additional scattering processes in the medium which leads to a significant enhancement in the DPR at lower values of invariant mass. The constituent quark mass evaluated using a 3-flavour Nambu--Jona-Lasinio model is also non-trivially affected by the CCP. These are found to result in a continuous dilepton production rate as a function of the invariant mass for higher values of temperature and baryonic chemical potential.		
\end{abstract}

	\maketitle
	
	\section{Introduction}
	The study of QCD vacuum structure under extreme conditions of temperature and/or baryon density is one of the main objectives of relativistic heavy ion collision (HIC) experiments at RHIC and LHC. 	It is well established that the infinite number of energy-degenerate different vacuum configurations of QCD at zero and low temperatures can be characterised by topologically non-trivial gauge configurations with a non-zero winding number~\cite{Shifman:1988zk}. These gluon configurations are called instantons which can invoke transition between two different vacua by means of crossing a potential barrier with a height of the order of the QCD scale $\Lambda_\text{QCD}$. This mechanism is known as instanton tunnelling~\cite{Belavin:1975fg,tHooft:1976rip,tHooft:1976snw}. However, at high temperatures, for example, in the quark gluon plasma (QGP) phase of HICs, a copious production of another kind of gluon configuration, called sphalerons, is expected~\cite{Manton:1983nd,Klinkhamer:1984di}. It is conjectured that the abundance of sphalerons can enhance the transition rate by crossing the barriers between different energy-degenerate vacua~\cite{Kuzmin:1985mm,Arnold:1987mh,Khlebnikov:1988sr,Arnold:1987zg}. The topologically non-trivial guage field configurations can switch the helicities of quarks while interacting with them. This in turn leads to the breaking of parity ($ P $) and charge-parity ($ CP $) symmetries by creating an asymmetry between left and right handed quarks via the axial anomaly of QCD~\cite{Adler:1969gk,Bell:1969ts}. Chirality imbalance can be produced locally as there is no direct observation	of the violation of $P$ and $CP$ in QCD globally~\cite{Adler:1969gk,Bell:1969ts,McLerran:1990de,Moore:2010jd}. This locally induced chirality imbalance is characterised by means of a chiral chemical potential (CCP) which basically represents the difference between the number of right and left-handed quarks.

%	
%	 in fact, invoke passing a potential barrier 	with a height of the order of the QCD scale $\Lambda_{QCD}$. These transition	rates are highly suppressed at low temperatre because of the height of the  potential barrier~\cite{Belavin:1975fg,tHooft:1976rip,tHooft:1976snw}. At high temperature, for example, in the quark gluon
%	plasma (QGP) phase, a copious production of another kind of gluon 
%	configurations, called sphalerons, is expected~\cite{Manton:1983nd,Klinkhamer:1984di,Kuzmin:1985mm,Arnold:1987mh,Khlebnikov:1988sr,Arnold:1987zg}. These kind of transitions can
%	lead to the breaking of parity ($P$) and charge-parity ($CP$) symmetry of
%	the high temperatre plasma through axial anomaly. As a result of this, 
%	 To characterize the locally
%	, a chiral chemical potential (CCP), is 
%	usually introduced, which leads to the difference between the number of
%	right- and left-handed quarks. 
	
	HICs with a non-zero impact parameter can give rise to very high magnetic fields of the order of few $ m_\pi^2 $~\cite{Kharzeev:2007jp,Skokov:2009qp}. Such high magnetic fields in presence of chirality imbalance can lead to a separation of positive and negative charges with respect to the reaction plane and induce a current along the magnetic field dubbed as chiral magnetic effect  (CME)~\cite{Fukushima:2008xe,Kharzeev:2007jp,Kharzeev:2009pj,Bali:2011qj}. Substantial efforts have been made to detect CME in HIC experiments at the RHIC at Brookhaven. Very recently, the STAR Collaboration has performed  an extensive analysis which has provided no indication of CME in HICs~\cite{STAR:2021mii}. As a consequence, new techniques for experimental determination of CME have been proposed~\cite{An:2021wof,Milton:2021wku}. 
	
	Since a local domain of chirality imbalance is expected to be produced in QGP, in addition to the CME there have been intense studies on the phase structure~\cite{Ruggieri:2016lrn,Ruggieri:2016asg,Ruggieri:2020qtq}, microscopic transport phenomena ~\cite{Vilenkin:1979ui,Vilenkin:1980fu,Fukushima:2008xe,Son:2009tf}, collective oscillations ~\cite{Akamatsu:2013pjd,Carignano:2018thu,Carignano:2021mrn}, fermion damping rate~\cite{Carignano:2019ivp} and collisional energy loss of fermions~\cite{Carignano:2021mrn} as well as properties of electromagnatic spectral function~\cite{Ghosh:2022xbf} in chirally imbalanced medium. Moreover, chirally asymmetric plasma is expected to be produced in the gap regions of the magnetospheres of pulsars and black holes~\cite{Gorbar:2021tnw} and other stellar astrophysical scenario~\cite{Charbonneau:2009ax,Akamatsu:2013pjd,Kaminski:2014jda,Yamamoto:2015gzz,Shovkovy:2021yyw}. Furthermore, it is worthwhile to mention that CME has indeed been observed in condensed matter systems particularly in 3D Dirac as well as Weyl semimetals~\cite{Li:2014bha,Li:2016vlc,Kharzeev:2013ffa,Kharzeev:2015znc,Huang:2015oca,Landsteiner:2016led,Gorbar:2017lnp,Joyce:1997uy,Tashiro:2012mf}. Thus the study of the properties of chirally imbalanced matter continues to be a matter of major topical interest.

	 It is well known that the hot and dense matter produced in HICs cools via rapid expansion under its own pressure passing through different stages of evolution. However, the whole process is very transient ($\sim$few fm/c) restricting the possibility of a direct observation. So to investigate microscopic as well as bulk properties of QGP one has to rely on indirect probes and observables~\cite{Wong:1995jf}. 
	% The electromagnetic spectral function (photon correlation function) is one %such tool which  plays a central role in the determination of various transport coefficients as well as in the production rates of dileptons and photons~\cite{Mallik:2016anp,Wong:1995jf,Alam:1999sc, Sarkar:2012ty} produced from HICs. 
	  Electromagnetic probes, photons and dileptons have long been used as reliable probes of HICs. Because they participate only in electromagnetic interaction, their mean free paths are much larger than the typical size  of the system. As a consequence, once produced they tend to leave the system without suffering further interactions, thus carrying unaltered information about the space-time region from where they are produced~\cite{McLerran:1984ay,Kajantie:1986dh,Weldon:1990iw,Alam:1996fd,Alam:1999sc,Rapp:1999ej,Aurenche:2000gf,Arnold:2001ms,Rapp:2009my,Chatterjee:2009rs,Sarkar:2012ty}. As the production rate of dileptons and photons is directly proportional to the electromagnetic spectral function, the study of its analytic properties  is of central importance~\cite{Mallik:2016anp,Wong:1995jf,Alam:1999sc, Sarkar:2012ty}. Such a study in chirally imbalanced hot and/or dense matter has recently been performed in~\cite{Ghosh:2022xbf} where a rich cut-structure was revealed. The step-like structure of the spectral function observed in certain invariant mass range were attributed to the  thresholds of Unitary and Landau cuts indicating additional scattering processes in the medium. We expect that such structures will have non-trivial effects on the dilepton spectrum.

	 The imaginary part of the electromagnetic current correlator containing the modified quark propagators in the presence of a hot and dense medium is the most important component in the evaluation of the dilepton production rate (DPR) which determines the thresholds as well as the intensity of emission of dileptons~\cite{Alam:1996fd,Alam:1999sc}. Thus it has a crucial dependence  in the value of quark mass. As the system cools, the quark condensate builds up due to the breaking of chiral symmetry which results in a large value of the quark mass ($ \sim $ few hundred MeV). The non-perturbative  nature of QCD at low energies severely hinders the theoretical analysis of these phenomena using first principle calculations.  As an alternative, we have used the Nambu--Jona-Lasinio (NJL) model~\cite{Nambu1,Nambu2}  which is built by respecting the global symmetries of QCD, most importantly the chiral symmetry~\cite{Klevansky,Buballa,Vogl}. This model has been very useful to probe the vacuum structure of QCD at arbitrary values of temperature, baryon chemical potential (BCP) and CCP~\cite{Ruggieri:2011xc,Fukushima:2010fe,Farias:2016let,Chao:2013qpa,Yu:2014sla,Yu:2015hym,Chaudhuri:2021lui}.

	 In this work, we shall evaluate the dilepton production rate (DPR) from a (locally) chirally imbalanced quark matter expected to be produced in relativistic HIC experiments. We use the 3-flavor NJL model to evaluate the constituent quark mass by solving the self-consistent gap equations. The quark mass for different flavours $M_f=M_f(T,\mu_B,\mu_5)$ will go as an input in the electromagnetic spectral function.
	 We have made use of the analytic structure of the in-medium spectral function to obtain the thresholds of dilepton production due to various scattering processes involving quarks for both zero and non-zero values of $ \mu_5 $. 
	 %\Fixme{We	propose that the dilepton production from chirally imbalanced medium	provides the possible platform to determine the experimental evidence for the existence of non-trivial topological gluon configurations, as well	as local $P$ and $CP$ violations}. 
	
	The paper is organized as follows. In the next section we discuss the formulation of the dilepton production rate at zero and finite value of the CCP. Next, we discuss in Section III the evaluation of the constituent quark mass from a 3-flavour NJL model. Section IV deals with the numerical results followed by a summary and discussion in Section V.
	
%~~~~~~~~~~~~~~~~~~~~~~~~~~~~~~~~~~~~~~~~~~~~~~~~~~~~~~~~~~~~~~~~~~~~~~~~~~~~~~~~~~~~~~~~~~~~~~~~~~~~~~~~~~~~	
	\section{DILEPTON PRODUCTION RATE} \label{sec.DPR}
	
\begin{figure}[h]
		\begin{center}
			\includegraphics[angle=0,scale=0.5]{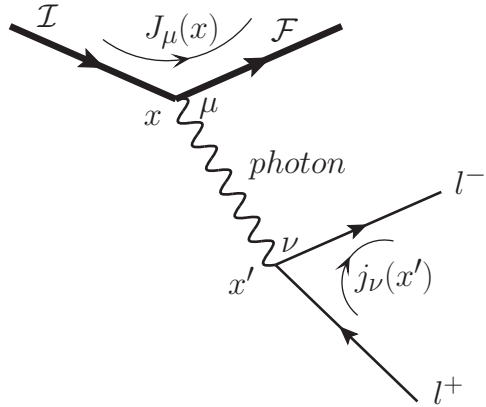}
		\end{center}
		\caption{Diagrammatic representation of the dilepton production amplitude.}
		\label{DL_diagram}
	\end{figure}

	In QGP, a quark can interact with an anti-quark to produce a virtual photon, which subsequently decays into a lepton $ l^+ $ and anti-lepton $ l^- $ pair (see Fig.~\ref{DL_diagram}).
	The dilepton production rate (DPR) from a hot and dense medium is already calculated in Refs.~\cite{McLerran:1984ay,Weldon:1990iw,Wong:1995jf,Mallik:2016anp,Bandyopadhyay:2017raf,Ghosh:2018xhh}. But, for the sake of completeness, we will briefly demonstrate few important steps. First, we consider an initial state  $\ket{\mcI}=\ket{I(p_I)}$ of a quark/antiquark with momentum $p_I$ moving towards to a final state $\ket{\mcF}=\ket{F(p_F),l^+(p_{l^+})l^-(p_{l^-})}$ consisting a quark/antiquark of momentum $p_F$ plus a pair of leptons of momenta $p_{l^+}$ and $p_{l^-}$ respectively. The probability amplitude for such a transition is  $|\expectationvalue{\mcF}{\mcSh}{\mcI}|^2$, where $\mcSh$ is the scattering matrix expressed as
	\begin{eqnarray}
	\mcSh = \mcT \TB{ \exp \FB{ i\int \scrL_\text{int}(x)d^4x } }
	\end{eqnarray}
	in which $\mcT$ is the time-ordering operator and 
	\begin{eqnarray}
	\scrL_\text{int}(x) = j^\mu(x)A_\mu(x)+J^\mu(x)A_\mu(x)
	\end{eqnarray}
	is the Lagrangian (density) for local interaction. Our choice of metric tensor is $g^\munu=\text{diag}(1,-1,-1,-1)$. In the above equation, the conserved vector currents corresponding to the leptons and quarks denoted by $j^\mu(x)$ and $J^\mu(x)$ respectively, are coupled to the photon field represented by  $A^\mu(x)$ . It can be shown that, the first non-trivial contribution to the above mentioned process comes from the second order term in the $ \mcSh $ matrix expansion and the expression of the squared amplitude $|\expectationvalue{\mcF}{\mcSh}{\mcI}|^2$ is given by,
	\begin{eqnarray}
	|\expectationvalue{\mcF}{\mcSh}{\mcI}|^2 = \int\!\!\!\int d^4x' d^4x e^{i(p_{l^+}+p_{l^-})\cdot x'}\frac{1}{(p_{l^+}+p_{l^-})^4}
	\ExpectationValue{l^+(p_{l^+})l^-(p_{l^-})}{j^\mu(0)}{0} \ExpectationValue{0}{j^{\nu\dagger}(0)}{l^+(p_{l^+})l^-(p_{l^-})} \nn \\
	\ExpectationValue{F(p_F)}{J_\mu(x')}{I(p_I)} \ExpectationValue{I(p_I)}{J_\nu^\dagger(0)}{F(p_F)}. \label{eq.sfi}
	\end{eqnarray}
	Now the dilepton multiplicity from thermal QGP medium is expressed as~\cite{Mallik:2016anp}
	\begin{eqnarray}
	N = \frac{1}{\mcZ} \sum_{\text{spins}}^{}\int\!\!\frac{d^3p_{l^+}}{(2\pi)^32p_{l^+}^0}\int\!\!\frac{d^3p_{l^-}}{(2\pi)^32p_{l^-}^0} \sum_{I,F}^{}
	\exp\FB{-\beta p_I^0} |\expectationvalue{F}{\mcSh}{I}|^2 \label{eq.DM}
	\end{eqnarray}
	where $\mcZ$ is the partition function of the system and the summation signifies  sum over all leptonic spin configurations. Using Eq.~\eqref{eq.sfi} in Eq.~\eqref{eq.DM} and after some simplifications, we arrive at~\cite{Ghosh:2018xhh}
	\begin{eqnarray}
	N = \int\!\! d^4x\int\!\!\! \frac{d^4q}{(2\pi)^4} e^{-\beta q^0}\frac{1}{q^4} \mtW_{+\munu}(q) \mtL_+^\munu(q) 
	\label{eq.dilepton.mult}
	\end{eqnarray}
	where,
	\begin{eqnarray}
	\mtW_+^\munu(q) &=& \int\!\! d^4x e^{iq\cdot x} \ensembleaverage{J^\mu(x)J^{\nu\dagger}(0)}, \label{eq.W+}\\
	\mtL_+^\munu(q) &=& \int\!\! d^4x e^{iq\cdot x} \ExpectationValue{0}{j^{\nu\dagger}(x)j^{\mu}(0)}{0}. \label{eq.L+}
	\end{eqnarray}
	Here $\ensembleaverage{...}$ represents the ensemble average and  $ q=(p_{l^+}+p_{l^-}) $ is the total momentum of the lepton-pair. So the dilepton production rate (DPR) becomes
	\begin{eqnarray}
	\text{DPR} = \frac{dN}{d^4xd^4q} = \frac{1}{(2\pi)^4} \frac{e^{-\beta q^0}}{q^4}  \mtW_{+\munu}(q) ~\mtL_+^\munu(q). \label{eq.DPR.1}
	\end{eqnarray}
	
	Now to calculate DPR in presence of a medium it is useful to write both the $\mtW_+^\munu(q)$ and $\mtL_+^\munu(q)$ in terms of time ordered correlation functions so that one can apply the real time formulation of finite temperature field theory~\cite{Mallik:2016anp,Ghosh:2018xhh,Ghosh:2020xwp,Chaudhuri:2021skc}. Thus we get
	\begin{eqnarray}
	\text{DPR} = \frac{dN}{d^4xd^4q} = \frac{1}{4\pi^4q^4} \FB{\frac{1}{e^{\beta q^0}+1}}\IM~ \mtW_{11}^\munu(q)\IM~ \mtL_\munu(q). \label{eq.DPR.2}
	\end{eqnarray}
	
	Till now we have not specified any explicit form of the currents $J^\mu(x)$ and $j^\mu(x)$ which are required to evaluate the quantities $\mtW_{11}^\munu(q)$ and $\mtL^\munu(q)$. In this work, we have considered
	\begin{eqnarray}
	J^\mu(x) &=& e\bar{{\rm q}}(x)\hat{Q}\gamma^\mu {\rm q}(x)~, \label{eq.J} \\
	j^\mu(x) &=& -e\psibar(x)\gamma^\mu\psi(x) \label{eq.j}
	\end{eqnarray}
	where, ${\rm q}=\FB{u~~d~~s}^T$ represents $ 3 $-flavor quark field multiplet with corresponding charge matrix $\hat{Q}= \text{diag}~ \FB{ \frac{2}{3}, -\frac{1}{3}, -\frac{1}{3}  } $, $\psi$ is the lepton field and $e>0$ is the absolute value of the electric charge of an electron. Using Eqs.~\eqref{eq.J} and \eqref{eq.j} it can be shown that
	\begin{eqnarray}
	\mtW_{11}^\munu(q) &=& i\int\!\!\!\frac{d^4k}{(2\pi)^4} \Tr_\text{d,f,c} 
	\TB{\gamma^\mu \hat{Q} S_{11}(p=q+k)\gamma^\nu \hat{Q} S_{11}(k)}~, \label{eq.W11.2}\\
	\mtL^\munu(q) &=& i e^2\int\!\!\!\frac{d^4k}{(2\pi)^4} \Tr_\text{d} \TB{\gamma^\nu S(p=q+k)\gamma^\mu S(k)} \label{eq.L.2}
	\end{eqnarray}
	where the trace over Dirac, flavor and color spaces are indicated by the subscript `d', `f', and `c' respectively; $S_{11}(p)$ is the 11-component of the real time quark propagator and $S(k)$ is the vacuum propagator for leptons with Feynman boundary condition which is expressed as 
	\begin{eqnarray}
	S(p) = \frac{-(\cancel{p}+m_L)}{p^2-m_L^2+i\varepsilon} \label{eq.lepton.propagator}
	\end{eqnarray}
	with $m_L$ being the mass of the lepton. 
	It is to be noted that, in Eq.~\eqref{eq.W11.2}, the quark propagator $S_{11}(p)$ is diagonal in both the flavor and color space i.e. 
	\begin{eqnarray}
	S_{11}(p) = \text{diag}\FB{S_{11}^u(p) , S_{11}^d(p), S_{11}^s(p)} \otimes 1_\text{color}.
	\label{eq.don}
	\end{eqnarray}

	Since both the currents $J^\mu(x)$ and $j^\mu(x)$ are conserved: $\del_\mu J^\mu(x)= 0 =\del_\mu j^\mu(x)$, consequently the matter tensor $\mtW_{11}^\munu(q)$ as well as the leptonic tensor $\mtL^\munu(q)$ are transverse to the momentum $q^\mu$ i.e.
	\begin{eqnarray}
	q_\mu ~\mtW_{11}^\munu(q) = 0 = q_\mu~ \mtL^\munu(q) . \label{eq.transversality}
	\end{eqnarray}
	These transversality conditions enforce the following Lorentz structure of $\mtL^\munu(q)$:
	\begin{eqnarray}
	\mtL^\munu(q) = \FB{g^\munu-\frac{q^\mu q^\nu}{q^2}}\FB{\frac{1}{3}g_{\rho\sigma} \mtL^{\rho\sigma}}. \label{eq.L.0}
	\end{eqnarray}
	Putting this back in Eq.~\eqref{eq.DPR.2}, and using the transversality condition for  matter tensor we get the DPR as
	\begin{eqnarray}
	\text{DPR} = \FB{\frac{dN}{d^4xd^4q}} = \frac{1}{12\pi^4q^4} \FB{\frac{1}{e^{q^0/T}+1}}
	g_\munu\IM ~\mtW_{11}^\munu(q) g_{\rho\sigma} \IM~ \mtL^{\rho\sigma}(q). \label{eq.DPR.3}
	\end{eqnarray}
	It is easy to check, by  substituting Eq.~\eqref{eq.lepton.propagator} into Eq.~\eqref{eq.L.2}, that 
	%	$g_{\rho\sigma} \IM \mtL^{\rho\sigma}(q)$ can be calculated in a straightforward way by  substituting Eq.~\eqref{eq.lepton.propagator} into Eq.~\eqref{eq.L.2} and performing some algebraic steps we get
	\begin{eqnarray}
	g_{\rho\sigma} \IM ~ \mtL^{\rho\sigma}(q) = \frac{-e^2}{4\pi}q^2\FB{1+\frac{2m_L^2}{q^2}}\sqrt{1-\frac{4m_L^2}{q^2}}\Theta\FB{q^2-4m_L^2}.
	\label{eq.L.3}
	\end{eqnarray}	
	The effects of finiteness of CCP will be encoded in the matter tensor part of Eq.~\eqref{eq.DPR.3}. Thus we shall consider the following two cases. 
%~~~~~~~~~~~~~~~~~~~~~~~~~~~~~~~~~~~~~~~~~~~~~~~~~~~~~~~~~~~~~~~~~~~~~~~~~~~~~~~~~~~~~~~~~~~~
	\subsection{DPR at $ \mu_5 = 0 $}\label{subsection_mu5_0}
	To evaluate matter part at finite temperature for vanishing values of CCP, one requires the 11-component of the real time thermal quark propagator. 
	In the spectral representation this can be expressed as~\cite{Mallik:2016anp}
	\begin{eqnarray}
	S_{11}^f(p) = \int_{-\infty}^{\infty} \frac{dp_0^\prime }{2 \pi} \sigma'^f ( p_0^\prime, \vec{p} ) \TB{\frac{1}{p_0^\prime -p_0 - i\varepsilon} - 2 \pi i \eta(p_0^\prime) \delta(p_0^\prime - p_0)}\label{eq.S11.T}
	\end{eqnarray}
	where, $ \sigma'^f ( p_0, \vec{p} ) = 2\pi ~\text{sgn} (p_0) \FB{\cancel{p} + M_f} \delta ( p^2 - M_f^2) $ is the fermionic spectral function with $M_f$ being the constituent mass of a quark with flavour $ f $ and $\eta (x)  $ is the distribution like function containing the true Fermi-Dirac thermal distributions $f^\pm(x)$ of the quarks given by
	\begin{eqnarray}
	 \eta(x)=\Theta(x)f^+(\MB{x})- \Theta(-x)f^-(\MB{x}) + \Theta(-x) ~~~~~\text{with}~~~~~  f^\pm(x) = \TB{\exp\FB{\frac{x\mp\mu_B/3}{T}}+1}^{-1} ~\label{eq.Dist_like_fn}
	\end{eqnarray}
	in which $\mu_B$ is the BCP. Now in the local rest frame (LRF) of the medium, using Eqs.~\eqref{eq.S11.T}, \eqref{eq.Dist_like_fn} and \eqref{eq.don} in Eq.~\eqref{eq.W11.2} one gets after some algebra
	\begin{eqnarray}
	g_\munu\IM ~\mtW_{11}^\munu(q) = N_c\sum_{f} e_f^2 \pi \int\frac{d^3k}{(2\pi)^3}&&\frac{1}{4\omega_k\omega_p}\big[
	\SB{1-f^-(\omega_k)-f^+(\omega_p)+2f^-(\omega_k)f^+(\omega_p)}\msN(k^0=-\omega_k)\delta(q^0-\omega_k-\omega_p) \nn \\
	&&+ \SB{1-f^+(\omega_k)-f^-(\omega_p)+2f^+(\omega_k)f^-(\omega_p)}\msN(k^0=\omega_k)\delta(q^0+\omega_k+\omega_p) \nn \\
	&&+ \SB{-f^-(\omega_k)-f^-(\omega_p)+2f^-(\omega_k)f^-(\omega_p)}\msN(k^0=-\omega_k)\delta(q^0-\omega_k+\omega_p) \nn \\
	&&+ \SB{-f^+(\omega_k)-f^+(\omega_p)+2f^+(\omega_k)f^+(\omega_p)}\msN(k^0=\omega_k)\delta(q^0+\omega_k-\omega_p) \big]
	\label{eq.W11.7}
	\end{eqnarray}
	where, $e_f=q_f e$, $\omega_k=\sqrt{\vec{k}^2+M_f^2}$, $\omega_p=\sqrt{\vec{p}^2+M_f^2}=\sqrt{(\vec{q}+\vec{k})^2+M_f^2}$ and $\msN (q,k)= 8(k^2+q\cdot k - 2M_f^2)$.  Note that, the first delta function appearing in  Eq.~\eqref{eq.W11.7}, termed as the unitary-I cut, represents to the contribution from quark-antiquark  annihilation to a positive energy time-like virtual photon (and the corresponding time reversed process where such a photon decays into a quark-antiquark pair). The delta function in the second term (the unitary-II cut) corresponds to similar process but the virtual photon is of negative energy.  The last two delta functions are called the Landau cuts, which are purely medium dependent contributions, stand for the scattering/emission processes such as absorption of a space-like virtual photon by a quark/antiquark  and the consequent time reversed processes. It can be easily checked that, the contributions from unitary-I and unitary-II cuts are non-zero in the kinematic regions $\sqrt{\vec{q}^2+4M_f^2}<q^0<\infty$ and $-\infty<q^0<-\sqrt{\vec{q}^2+4M_f^2}$ respectively.
	\begin{figure}[h]
		\begin{center}
			\includegraphics[angle=0,scale=0.5]{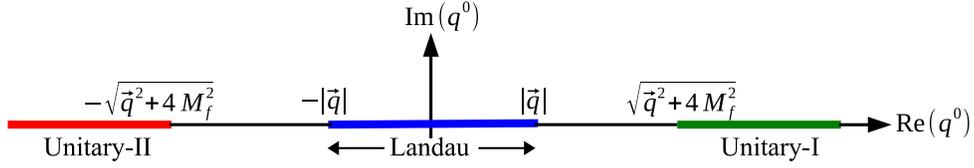}
		\end{center}
		\caption{(Color Online) The branch cuts of $\mtW_{11}^\munu(q)$ in the complex $q^0$ plane for a given $\vec{q}$. Kinematic domain for the physical dileptons production defined in terms of $q^0>0$ and $q^2>0$ corresponds to the green region.}
		\label{fig.analytic0}
	\end{figure}
	 On the other hand, the kinematic domain for both the Landau cuts is in the space-like region $|q^0|<|\vec{q}|$. This cut structure of $ {\rm Im} \mathtt{W}^{\munu}_{11} $ in the complex $ q^0 $-plane is depicted in Fig.~\ref{fig.analytic0}.	 Since, we are interested in the physical dileptons with positive energy and time-like four momentum i.e $ q^0 >0 $ and $q^2>0$, it follows that, only the unitary-I cut contributions are kinematically allowed. It should be noted that, the kinematic domains are directly related to constituent quark mass $ M_f $ and hence can be different for different flavors~\cite{Ghosh:2018xhh,Ghosh:2020xwp,Chaudhuri:2021skc}.
	Now, with the physical restrictions previously mentioned, the $d^3k = \kvec^2d\kvec d(\cos\theta) d\phi$ integral of Eq.~\eqref{eq.W11.7} can be evaluated analytically (the $d(\cos\theta)$ integral has been performed using the Dirac delta functions present in the integrand and the $d\phi$ integral trivially gives a factor of $2\pi$). Using this result in Eq.~\eqref{eq.DPR.3}, we finally arrive at the following analytical expression for the DPR at vanishing CCP~\cite{Ghosh:2018xhh,Greiner:2010zg}
	\begin{eqnarray}
	\text{DPR}_{\mu_5=0} = \FB{\frac{dN}{d^4xd^4q}}_{\mu_5=0} &=&  \sum_f \Theta\FB{q^2-4m_L^2} \Theta\FB{q^2-4M_f^2}  N_c\frac{ e^2 e_f^2}{ 192\pi^6} \frac{1}{\beta|\vec{q}|} \FB{\frac{1}{e^{\beta q^0}-1}} \nn \\ 
	&& ~~~\times \FB{1+\frac{2m_L^2}{q^2}}\sqrt{1-\frac{4m_L^2}{q^2}}
	\FB{1+\frac{2M_f^2}{q^2}} \ln \TB{\SB{\frac{e^{\beta(q^0+q_-)}+1}{e^{\beta (q^0+q_+)}+1}}\FB{\frac{e^{\beta q_+}+1}{e^{\beta q_-}+1}}}
	\label{eq.DPR.4}
	\end{eqnarray}
	where $q_\pm = -\frac{1}{2}\TB{q^0 \pm |\vec{q}|\sqrt{1-\frac{4M_f^2}{q^2}}} + \mu_B/3$.
The step functions in the above expression restrict the production of dileptons  with invariant mass $ q^2<4m_L^2 $ and $ q^2<4M_f^2  $.
%%%===================================================================

\subsection{DPR at $ \mu_5 \ne 0 $} \label{subsection_mu5_ne_0}
\begin{figure}[h]
	\begin{center}
		\includegraphics[angle=0,scale=0.5]{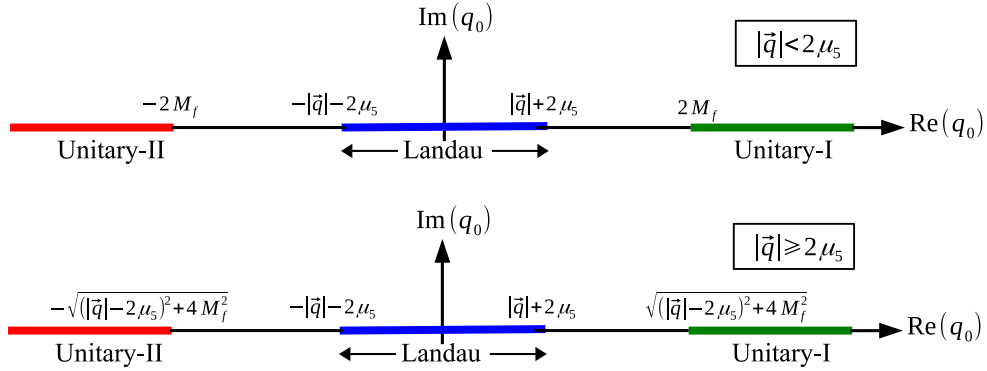} 
	\end{center}
	\caption{(Color Online) The branch cuts of the self energy in the complex $q_0$ plane for a given $\qvec$ when (upper panel)  $\qvec<2\mu_5$ (lower panel)  $\qvec\ge2\mu_5$. Kinematic domain for the physical dileptons production defined in terms of $q^0>0$ and $q^2>0$ corresponds to the green region and some portion of the blue region.}
	\label{Fig_ana_struc_mu5}
\end{figure}

This subsection is devoted to the evaluation of $ g_\munu\IM ~\mtW_{11}^\munu(q) $ in a medium in presence of chiral imbalance. Here, one needs the 11-component of the real time thermal quark propagator with finite values of CCP. In the spectral representation the later can be expressed as~\cite{Mallik:2016anp,Ghosh:2022xbf}
	\begin{eqnarray}
S^f_{11}(p) &=& \int_{-\infty}^{\infty} \frac{dp_0^\prime }{2 \pi} \sigma^f ( p_0^\prime, \vec{p} ) \TB{\frac{1}{p_0^\prime -p_0 - i\varepsilon} - 2 \pi i \eta(p_0^\prime) \delta(p_0^\prime - p_0)} \nn \\
&=& \int_{-\infty}^{\infty} \frac{dp_0^\prime }{2 \pi} \sigma^f ( p_0^\prime, \vec{p} ) \TB{ \mcP \FB{\frac{1}{p^\prime_0 - p_0}}  - 2 \pi i \FB{\eta(p_0^\prime) - \frac{1}{2}} \delta (p_0^\prime - p_0) }
\label{eq.S11.T.mu5}
\end{eqnarray}
where the fermionic spectral function $\sigma^f (p_0, \vec{p})$ in presence of CCP is
\begin{eqnarray}
\sigma^f (p_0, \vec{p}) &=& 2 \pi~ {\rm sgn } (p_0) \frac{\scrD (p;M_f)}{{\omega_p^+}^2 - {\omega_p^-}^2} \TB{ \delta (p^2 - {\omega_p^+}^2) - \delta (p^2 - {\omega_p^-}^2) },
\end{eqnarray}
in which, $\omega_p^r = \sqrt{(|\vec{p}|+r\mu_5)^2+M_f^2}$ with $ r $ corresponds to the helicity of the propagating Fermion and $\mathscr{D}(p;M_f)$ contains the complicated Dirac structure as
\begin{eqnarray}
\mathscr{D}(p;M_f) &=& \sum_{j \in \{\pm\}} \mathscr{P}_j \TB{ p_{-j}^2\cancel{p}_j - M_f^2 \cancel{p}_{-j} + M_f (p_j\cdot p_{-j}-M_f^2) 
	+ i M_f \sigma_\munu p_j^\mu p_{-j}^\nu  }
\label{D}.
\end{eqnarray}
In the above equation, $ \mathscr{P}_j = \frac{1}{2}(1+j\gamma^5)$, $p_j^\mu \equiv(p^0+j\mu_5,\vec{p}) $ and $ \eta(x) $ is already defined in Eq.~\eqref{eq.Dist_like_fn}. Using this we get 
\begin{eqnarray}
\mtW_{11}^\munu &=& -i \sum_f e_f^2 \kfourint{k} \int_{-\infty}^{\infty} \frac{dk_0^\prime }{2 \pi}
\int_{-\infty}^{\infty} \frac{dp_0^\prime }{2 \pi}~ \dfrac{\mathscr{N}^\munu (k,q)~{\rm sgn} (k_0^\prime)~{\rm sgn} (p_0^\prime) }{\fb{{\omega_k^+}^2 - {\omega_k^-}^2} \fb{{\omega_p^+}^2 - {\omega_p^-}^2}} \SB{ \delta (k^2 - {\omega_k^+}^2) - \delta (k^2 - {\omega_k^-}^2) } \nn \\ 
&& \times \SB{ \delta (p^2 - {\omega_p^+}^2) - \delta (p^2 - {\omega_p^-}^2) }~ \TB{ \dfrac{1 - \eta(k_0^\prime)}{k_0^\prime - k_0 - i\varepsilon} + \dfrac{\eta(k_0^\prime)}{k_0^\prime - k_0 + i\varepsilon} }~\TB{ \dfrac{1 - \eta(p_0^\prime)}{p_0^\prime - p_0 - i\varepsilon} + \dfrac{\eta(p_0^\prime)}{p_0^\prime - p_0 + i\varepsilon} }
\end{eqnarray}
with $ \mathscr{N}^\munu (k,q) = \Tr_\td \TB{ \gamma^\nu \scrD(p)\gamma^\mu \scrD(k) } $. Now contracting $ \mtW_{11}^\munu $ with the metric tensor and concentrating on the imaginary part we get
\begin{eqnarray}
g_\munu\IM~\mtW^\munu_{11} (q_0,\vec{q}) &=& -N_c\sum_{f  }e_f^2 \pi \int \frac{d^3k}{(2\pi)^3} \sum_{r \in \{\pm\}} \sum_{s \in \{\pm\}} 
\frac{1}{16rs\mu_5^2 \pvec\kvec} \frac{1}{4\wkr\wps} \nn \\
&& \times \Big[ \mathscr{N}(k^0=-\wkr) \SB{1-f_-(\wkr)-f_+(\wps)+2f_-(\wkr)f_+(\wps)}\delta(q_0-\wkr-\wps) \nn \\
&& ~+~ \mathscr{N}(k^0=\wkr) \SB{1-f_+(\wkr)-f_-(\wps)+2f_+(\wkr)f_-(\wps)}\delta(q_0+\wkr+\wps) \nn \\
&& ~+~ \mathscr{N}(k^0=\wkr) \SB{-f_+(\wkr)-f_+(\wps)+2f_+(\wkr)f_+(\wps)}\delta(q_0+\wkr-\wps) \nn \\
&& ~+~ \mathscr{N}(k^0=-\wkr) \SB{-f_-(\wkr)-f_-(\wps)+2f_-(\wkr)f_-(\wps)}\delta(q_0-\wkr+\wps) \Big]
\label{gImW11.1}
\end{eqnarray}
where,
%\begin{eqnarray}
%\mathscr{N} \FB{q,k} &=& g_\munu \mathscr{N}^\munu = 8 ~\Big[~2 M_f^6 + 2 k_0^2 M_f^2 \mu_5^2 + 3 M_f^4 \mu_5^2 + 2 k_0^2 \mu_5^4 - \mu_5^6 + 
%2 \mu_5^2 \FB{M_f^2 + \mu_5^2} \FB{k_0 + q_0}^2 - 2 M_f^4 \FB{k^2 + 2 k\cdot q + q^2} \nn \\ && \hspace{-0.5IN} 
%+ 2 k_0^2 \mu_5^2 \FB{k^2 + 2 k\cdot q + q^2} - 3 M_f^2 \mu_5^2 \FB{k^2 + 2 k\cdot q + q^2} - 
%\mu_5^4 \FB{k^2 + 2 k\cdot q + q^2} + 
%2 k_0 \mu_5^2 \FB{k_0 + q_0} \FB{k^2 + 2 k\cdot q - 2 M_f^2 + q^2} \nn \\ &&\hspace{-0.5IN}   + \FB{k^2 + 
%	k\cdot q} \SB{-\FB{M_f^2 - \mu_5^2}^2 - 
%	4 k_0 \mu_5^2 \FB{k_0 + q_0} + \FB{M_f - \mu_5} \FB{M_f + \mu_5} \FB{k^2 + 2 k\cdot q + q^2}} - 
%k^2 \SB{\FB{M_f^2 + \mu_5^2} \FB{2 M_f^2 + \mu_5^2}\nn \right.\\ && \left.   - 
%	2 \mu_5^2 \FB{k_0 + q_0} \FB{2 k_0 + q_0} + \FB{-2 M_f^2 + \mu_5^2} \FB{k^2 + 2 k\cdot q + 
%		q^2} + \FB{k^2 + k\cdot q} \FB{k^2 + 2 k\cdot q - M_f^2 + \mu_5^2 + q^2}}~\Big]~.
%\end{eqnarray}
%
\begin{eqnarray}
\mathscr{N} \FB{q,k} = g_\munu \mathscr{N}^\munu &=& 
16 M_f^6 -4 M_f^4 \big\{4 (k_+\cdot k_-)+(k_+\cdot p_+)+(k_-\cdot p_-)+4 (p_+\cdot p_-)\big\} \nn \\
&&~ +4 M_f^2 \big\{(k_+\cdot p_-) (k_-^2+p_+^2)+(k_-\cdot p_+) (k_+^2+p_-^2)+4 (k_+\cdot k_-) (p_+\cdot p_-) \big\} \nn \\
&&~ -4 \big\{ k_-^2 (k_+\cdot p_+) p_-^2+k_+^2 (k_-\cdot p_-) p_+^2 \big\}.
\end{eqnarray}
Here, some discussions related to the analytic structure of $ \IM~ \mtW_{11} $ in the complex $ q_0 $ plane are in order. It can be seen that, in presence of finite $ \mu_5 $, the imaginary part of the matter tensor consists of sixteen Dirac delta functions which leads to several branch cuts in the complex $q_0$ plane. The terms containing $\delta(q_0-\wkr-\wps)$ and $\delta(q_0+\wkr+\wps)$ are referred to as Unitary-I and Unitary-II cuts respectively as already mentioned in the previous section. However, in non-zero CCP case each of the Unitary cut consists of further sub-cuts owing to the different helicities $ (r,s) $. These different cuts correspond to different physical processes. For example, the Unitary-I (Unitary-II) cuts represent the decay of a virtual photon having positive (negative) energy to real quark-antiquark pair (and the corresponding time reversed process). The terms with the remaining two delta functions, i.e. $\delta(q_0+\wkr-\wps)$ and $\delta(q_0-\wkr+\wps)$, are called Landau-I and Landau-II cuts which also contain four sub-cuts corresponding to distinct helicities. As already mentioned, Landau cuts stand for the emission(absorption) processes in which a real quark/antiquark in the thermal medium emits (absorbs) a virtual photon. The detailed analysis to find out respective kinematic domains such that the imaginary part of $ ~\mtW_{11} $ receives non-trivial contributions from the sixteen different delta function has been done in~\cite{Ghosh:2022xbf}. Here we only quote the final result in tabular form in Eq.~\eqref{Table_Kinematic_Region}.
\begin{eqnarray}
\begin{tabular}{|c|c|}
\hline 
Cuts & \makecell[c]{ ~ \\ Kinematic Regions \\ ~ } \\
\hline \hline
Unitary-I & \makecell[c]{ ~ \\ $ 2M_f \le q_0 < \infty$ for $\qvec < 2\mu_5$ \\
	$ \sqrt{(\qvec-2\mu_5)^2+4M_f^2} \le q_0 < \infty$  for $\qvec \ge 2\mu_5$\\ ~ }\\
\hline
Unitary-II & \makecell[c]{ ~ \\ $ -\infty < q_0 \le -2M_f $ for $\qvec < 2\mu_5$ \\
	$ -\infty < q_0 \le -\sqrt{(\qvec-2\mu_5)^2+4M_f^2} $  for $\qvec \ge 2\mu_5$\\ ~ }\\
\hline
Landau-I \& Landau-II & \makecell[c]{ ~ \\ $ -\qvec-2\mu_5 \le q_0 \le \qvec+2\mu_5 $ \\ ~ }\\
\hline
\end{tabular}
\label{Table_Kinematic_Region}~.
\end{eqnarray}
The pictorial representation of the complex analytical structure of $ \IM ~\mtW^\munu_{11} $ expressed in Eq.~\eqref{Table_Kinematic_Region} is shown in Fig.~\ref{Fig_ana_struc_mu5}. Since we are interested in physical dileptons, we will restrict ourselves to the time-like kinematic domains with $q_0>0$ and $q^2>0$. Form Fig.~\ref{Fig_ana_struc_mu5} it is evident, in addition to the Unitary-I cut, some portion of the Landau cuts $\qvec < q_0 < 2\mu_5$ will also contribute to the $g_{\munu} \IM ~\mtW^\munu_{11} $ and hence to the DPR.  This contribution at lower invariant mass region is purely a finite CCP effect. Moreover, notice that, the thresholds of the Unitary cuts strongly depend on $\mu_5$ and $M_f$. Consequently, for $\mu_5 \ge \qvec/2$, a positive energy photon with $q^2 \ge 4(M_f^2-\mu_5^2)$ can in principle decay into a real quark-antiquark pair below the usual threshold of pair production $q^2 \ge 4M_f^2$; even for a space-like photon. Finally, for sufficiently high $\mu_5$, the forbidden gap between the Unitary and Landau cut contributions will vanish irrespective of the value of $M_f$ which enables the production of a continuous spectrum of dileptons through out the whole range of invariant mass, which is again possible only in a chirally asymmetric medium. 

Now, substituting Eqs.~\eqref{eq.L.3} and ~\eqref{gImW11.1} in Eq.~\eqref{eq.DPR.3} we get
%\begin{eqnarray}
%	\text{DPR}_{\mu_5\ne 0} &=& \FB{\frac{dN}{d^4xd^4q}}_{\mu_5\ne 0} = \frac{e^2q^2}{4\pi}\FB{1+\frac{2m_L^2}{q^2}}\sqrt{1-\frac{4m_L^2}{q^2}}\Theta\FB{q^2-4m_L^2}N_c \sum_{f  } e_f^2 \pi \int \frac{d^3k}{(2\pi)^3}\nn \\
%	&& \hspace{-0.7IN} \sum_{r \in \{\pm\}} \sum_{s \in \{\pm\}} 
%	\frac{1}{16rs\mu_5^2 \pvec\kvec} \frac{1}{4\wkr\wps}  \Big[ \mathscr{N}(k^0=-\wkr) \SB{1-f_-(\wkr)-f_+(\wps)+2f_-(\wkr)f_+(\wps)}\delta(q_0-\wkr-\wps) \nn \\
%	&& ~~~~~~~~~~~+~ \mathscr{N}(k^0=\wkr) \SB{1-f_+(\wkr)-f_-(\wps)+2f_+(\wkr)f_-(\wps)}\delta(q_0+\wkr+\wps) \nn \\
%	&& ~~~~~~~~~~~+~ \mathscr{N}(k^0=\wkr) \SB{-f_+(\wkr)-f_+(\wps)+2f_+(\wkr)f_+(\wps)}\delta(q_0+\wkr-\wps) \nn \\
%	&& ~~~~~~~~~~~+~ \mathscr{N}(k^0=-\wkr) \SB{-f_-(\wkr)-f_-(\wps)+2f_-(\wkr)f_-(\wps)}\delta(q_0-\wkr+\wps) \Big]
%	\label{DPR.mu5}
%\end{eqnarray}
%
\begin{eqnarray}
\text{DPR}_{\mu_5\ne 0} = \FB{\frac{dN}{d^4xd^4q}}_{\mu_5\ne 0} &=& 
\frac{e^2q^2}{4\pi}\FB{1+\frac{2m_L^2}{q^2}}\sqrt{1-\frac{4m_L^2}{q^2}}\Theta\FB{q^2-4m_L^2}N_c \sum_{f  } e_f^2 \pi \int\!\! \frac{d^3k}{(2\pi)^3}\sum_{r \in \{\pm\}} \sum_{s \in \{\pm\}}
\frac{1}{16rs\mu_5^2 \pvec\kvec} \nn \\
&& \times~ \frac{1}{4\wkr\wps}  \Big[ \mathscr{N}(k^0=-\wkr) \SB{1-f_-(\wkr)-f_+(\wps)+2f_-(\wkr)f_+(\wps)}\delta(q_0-\wkr-\wps) \nn \\
&& +~ \mathscr{N}(k^0=\wkr) \SB{1-f_+(\wkr)-f_-(\wps)+2f_+(\wkr)f_-(\wps)}\delta(q_0+\wkr+\wps) \nn \\
&& +~ \mathscr{N}(k^0=\wkr) \SB{-f_+(\wkr)-f_+(\wps)+2f_+(\wkr)f_+(\wps)}\delta(q_0+\wkr-\wps) \nn \\
&& +~ \mathscr{N}(k^0=-\wkr) \SB{-f_-(\wkr)-f_-(\wps)+2f_-(\wkr)f_-(\wps)}\delta(q_0-\wkr+\wps) \Big]
\label{DPR.mu5}
\end{eqnarray}
In Eq.~\eqref{DPR.mu5}, the angular $d(\cos\theta)$ integral is performed using the Dirac delta functions present in the integrand and the azimuthal $d\phi$ integral gives a factor of $2\pi$. However, unlike the vanishing CCP case, here the analytical evaluation of the remaining $d\kvec$ integral becomes cumbersome and hence will be evaluated numerically to obtain the DPR. Note that, while calculating DPR the constituent quark mass ($ M_f $) for different flavour, which depends on the external parameters, such as, temperature, BCP and CCP, are required. Here NJL model is used to calculate $ M_f $ in different physical conditions.

%~~~~~~~~~~~~~~~~~~~~~~~~~~~~~~~~~~~~~~~~~~~~~~~~~~~~~~~~~~~~~~~~~~~~~~~~~~~~~~~~~~~~~~~~~~~~~~~~~~~~~~~~~~~~~
\section{THE CONSTITUENT QUARK MASS using 3-flavor Nambu-Jona--Lasinio MODEL} \label{sec.NJL}
In this section we briefly outline few important steps to calculate constituent quark mass using NJL model. The Lagrangian for the 3-flavor gauged NJL model is given by
\begin{eqnarray}
\scrL &=& \overline{\text{q}}(x)\FB{i\fsl{\partial}-eQ\fsl{A}-\hat{m} + \gamma^0 \mu_q +\gamma^0 \gamma^5 \mu_5 }\text{q}(x)  
+ G_S \sum_{a=0}^{8}\SB{ \FB{\overline{\text{q}}(x) \lambda^a \ \text{q} (x)}^2 + \FB{ \overline{\text{q}}(x) i\gamma_5\lambda^a \text{q}(x)}^2} \nn \\&& \hspace{0.92IN}-~ K \TB{\det \overline{\text{q}}\FB{1 + \gamma_5} \text{q} + \det \overline{\text{q}}\FB{1 - \gamma_5} \text{q}} . \label{3NJL_lagrangian}
\end{eqnarray}
In the above expression, $ \text{q} = (u~ d~ s)^T $ is the quark field multiplet with three flavors ($ N_f = 3 $) and three colors ($ N_c = 3 $) (flavour and colour indices are suppressed). $ \hat{m}= \text{diag}(m_u , m_d , m_s ) $ is the current quark mass matrix. $ \lambda^a $s are the Gell-Mann matrices corresponding to the flavor $ SU_f(3) $. The isospin symmetry on the Lagrangian level is assumed, i.e., $ m_u = m_d = m_0 $ , while $ SU_f(3) $-symmetry is explicitly broken, so that $ m_s \ne m_0 $.  $ G_S $ is the the scalar  coupling  strength and the $ K $ term represents the six-point Kobayashi-Maskawa-t’Hooft (KMT) interaction which is responsible for breaking of the axial $ U(1)_A $ symmetry~\cite{Klevansky}. Using mean field approximation one arrives at the following gap equations: 
\begin{eqnarray}
M_u &=& m_u - 2 G_s \AB{\overline{\text{q}}\text{q} }_u + 2K \AB{\overline{\text{q}}\text{q} }_d \AB{\overline{\text{q}}\text{q} }_s  \label{Gap_M31}~, \\
M_d &=& m_d - 2G_s \AB{\overline{\text{q}}\text{q} }_d + 2K \AB{\overline{\text{q}}\text{q} }_s \AB{\overline{\text{q}}\text{q} }_u  \label{Gap_M32}~, \\
M_s &=& m_s - 2G_s \AB{\overline{\text{q}}\text{q} }_s + 2K \AB{\overline{\text{q}}\text{q} }_u \AB{\overline{\text{q}}\text{q} }_d  \label{Gap_M33}~,
\end{eqnarray}
where
\begin{equation}\label{text{q}bartext{q}}
\AB{\overline{\text{q}}\text{q} }_f = - N_c M_f \sum_r \kthreeint{p} \frac{1}{\wpr} \SB{ 1 - f_+ (\wpr) -f_- (\wpr) }
\end{equation}
The self-consistent solution of Eqs.~\eqref{Gap_M31}--\eqref{Gap_M33} results in $ T $ and/or $ \mu_B $ dependence of $ M_u, M_d, M_s  $ for different values of CCP. Notice that, the medium independent integral Eq.~\eqref{text{q}bartext{q}} is ultraviolet divergent. Since the NJL Lagragian is known to be non-renormalizable owing to the point-like interaction between the quarks~\cite{Klevansky}, one has to specify a proper regularization scheme.

To avoid a cutoff artifact, several smooth regularization procedure has been used in the literature  by introducing a form factor $ f_\Lambda $ in the diverging vacuum integrals. One can choose different functional form of this $ f_\Lambda $, such as, the Lorentzian type form factors~\cite{Fukushima:2010fe,Gatto:2010qs,Gatto:2010pt,Gatto:2012sp, Ghosh:2022xbf,Yu:2014sla} and Woods-Saxon type form factors~\cite{Fayazbakhsh:2010bh,Fayazbakhsh:2012vr,Fayazbakhsh:2014mca}. In this work, we have used first kind of smoothing function by introducing a multiplicative form factor~\cite{Fukushima:2010fe,Yu:2014xoa,Ghosh:2022xbf} 
\begin{equation}
f_\Lambda (p) = \sqrt{\frac{\Lambda^{2N_\Lambda}}{\Lambda^{2N_\Lambda} + {\pvec}^{2N_\Lambda}}}
\label{FormFactor}
\end{equation}
In the limit $ N_\Lambda \rightarrow \infty $ the form factor is reduced to the sharp cutoff function  $ \Theta (\Lambda - \MB{\vec{p}}) $ which infer that for larger values of $ N_\Lambda  $, the cutoff artifacts are expected to increase. On the other hand, for small $ N_\Lambda $ values, it is impossible to fit different phenomenological values, such as,  pion decay constant, the vacuum value of chiral condensate etc~\cite{Gatto:2010pt}. Here, we have taken  $ N_\Lambda =10 $ for numerical convenience.
The other model parameters are given in Table~1.
\begin{table}[h]
	\begin{center}
		\caption{Parameter set for $ 3 $-flavor model}
		\begin{tabular} { p{3cm}p{2cm}p{2cm}p{1.5cm}p{1.5cm} }
			\hline \hline
			\hspace{0.03IN}$ m_u = m_d $ (MeV) & $ m_s  $ (MeV)&  \hspace{0.1IN}$ \Lambda $ (MeV) &  \hspace{0.07IN}$ G_s\Lambda^2  $  & \hspace{0.08IN}$ K\Lambda^5 $ \\ 
			\hline
			\vspace{0.032IN} \hspace{0.2IN}$ 5.1 $ \vspace{0.032IN}& \vspace{0.032IN} \hspace{0.109IN}$133$ & \vspace{0.032IN} \hspace{0.15IN}$604.5$ & \vspace{0.032IN} \hspace{0.07IN}$ 3.25 $ &\vspace{0.032IN} \hspace{0.0IN} $ 10.58 $ \\
			\hline
		\end{tabular} \label{Table_3flv}
	\end{center}
\end{table}

These parameters are determined by fitting $  f_\pi $ , $ m_\pi $ ,$  m_K $ , and $ m_{\eta^\prime} $ to their phenomenological values~\cite{Yu:2014xoa}.

\section{Numerical Results}
	%++++++++++++++++++++++++++++++++++++++++++++++++++++++++++++++++++++++++++++++	
	\begin{figure}[h]
		\begin{center}
			\includegraphics[angle=-90,scale=0.4]{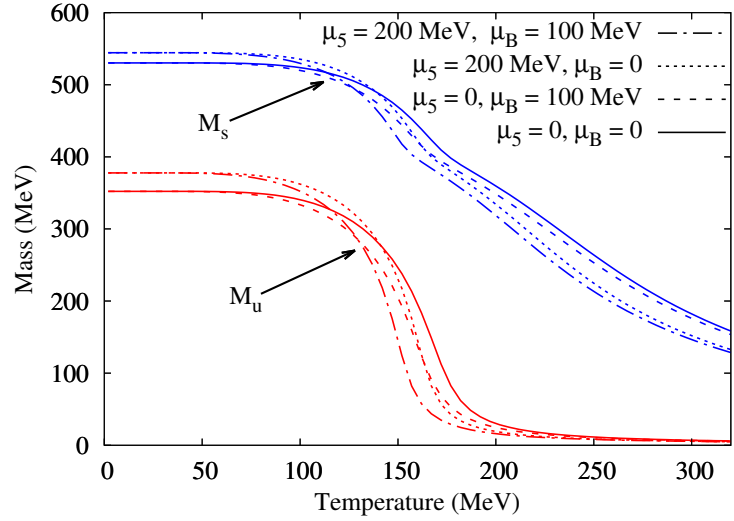}
		\end{center}
		\caption{(Color Online) Constituent quark mass as a function of temperature for different values of $ \mu_B $  and $ \mu_5 $. $M_u$ and $M_s$ are represented respectively by the red and blue curves.}
		\label{Fig_Mass}
	\end{figure}
In this section, we present the numerical results for constituent quark masses of  different flavors which are the main inputs while calculating the DPR from a  chirally asymmetric hot and dense medium. In Fig.~\ref{Fig_Mass}, we have shown the variation of constituent mass of 'up' and 'strange' quarks as a function of temperature for different values of BCP and CCP which are evaluated by solving Eqs.~\eqref{Gap_M31}-\eqref{Gap_M33} self-consistently. One can observe that, the spontaneous breaking of the chiral symmetry at small values of temperature results in large values of constituent mass for both `up' and `strange' quarks owing  to the large values of quark condensate for all the cases. Now, as the temperature is increased, the constituent mass of the low lying quarks remains constant upto a certain value of temperature, then falls off sharply in a small range of temperature and finally becomes nearly equal to the bare masses of the quarks at high $ T $ values representing the pseudo-chiral phase transition due to the (partial) restoration of the chiral symmetry. However, the strange quark mass decreases smoothly when compared to that of the up quark and it can be seen that even at $ T\sim 250  $~MeV the  s-quark mass is  still substantially higher than its current mass. 
The explanation of this behaviour comes from Eq.~\eqref{text{q}bartext{q}}. As the current quark mass becomes large, the  excitation probability of the quark-antiquark pair becomes thermally suppressed, which	leads to a  small $ T $-dependence of $ \AB{\bar{s}s} $ when compared to the low lying quarks~\cite{Hatsuda:1987pc}. Now at high temperature as the condensates of $ u $ and $ d $-quarks melt, the third term of Eq.~\eqref{Gap_M33} becomes negligible and the constituent mass of strange quark is solely determined by the $ \AB{\bar{s}s} $ condensate resulting in a smooth variation of $ M_s $. This indicates that $ SU(3)_f $ symmetry is not even a good approximate symmetry at temperatures larger than 200 MeV which may be due to the fact that the restoration of the chiral symmetry in the different quark sectors is achieved quite differently~\cite{Hatsuda:1994pi}.
Now for finite values of BCP, it is seen that the qualitative behaviour of the $ T $-dependence of the u and s-quark masses remain same, although  the transition temperature is found to decrease thus mimicking the conjectured QCD phase diagram. In presence of finite CCP, the constituent mass of both u and s-quarks are found to  increase in the low temperature region indicating an enhancement in the magnitude of the quark condensate at small values of $ T $. However, the transition from chirality broken to the restored phase  occurs at relatively smaller values of temperature which shows that, at high values of temperature, the formation of quark condensate is hindered by the presence of finite $ \mu_5 $, an exact opposite result compared to the  low temperature values. These phenomena can be termed as `chiral catalysis' and `inverse chiral catalysis' respectively~\cite{Ghosh:2022xbf}.

	\begin{figure}[h]
		\includegraphics[angle = -90, scale = 0.233]{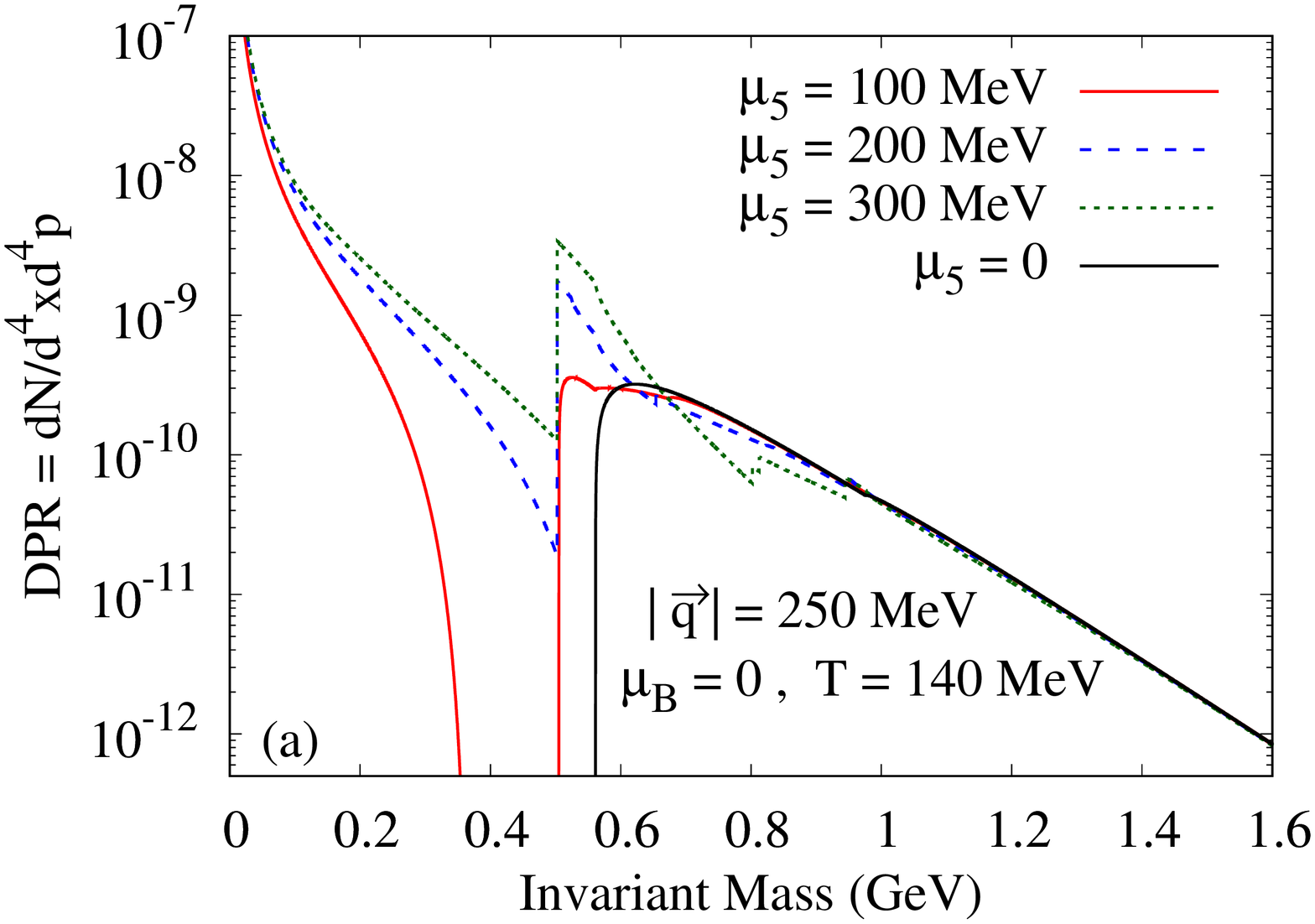}
		\includegraphics[angle = -90, scale = 0.233]{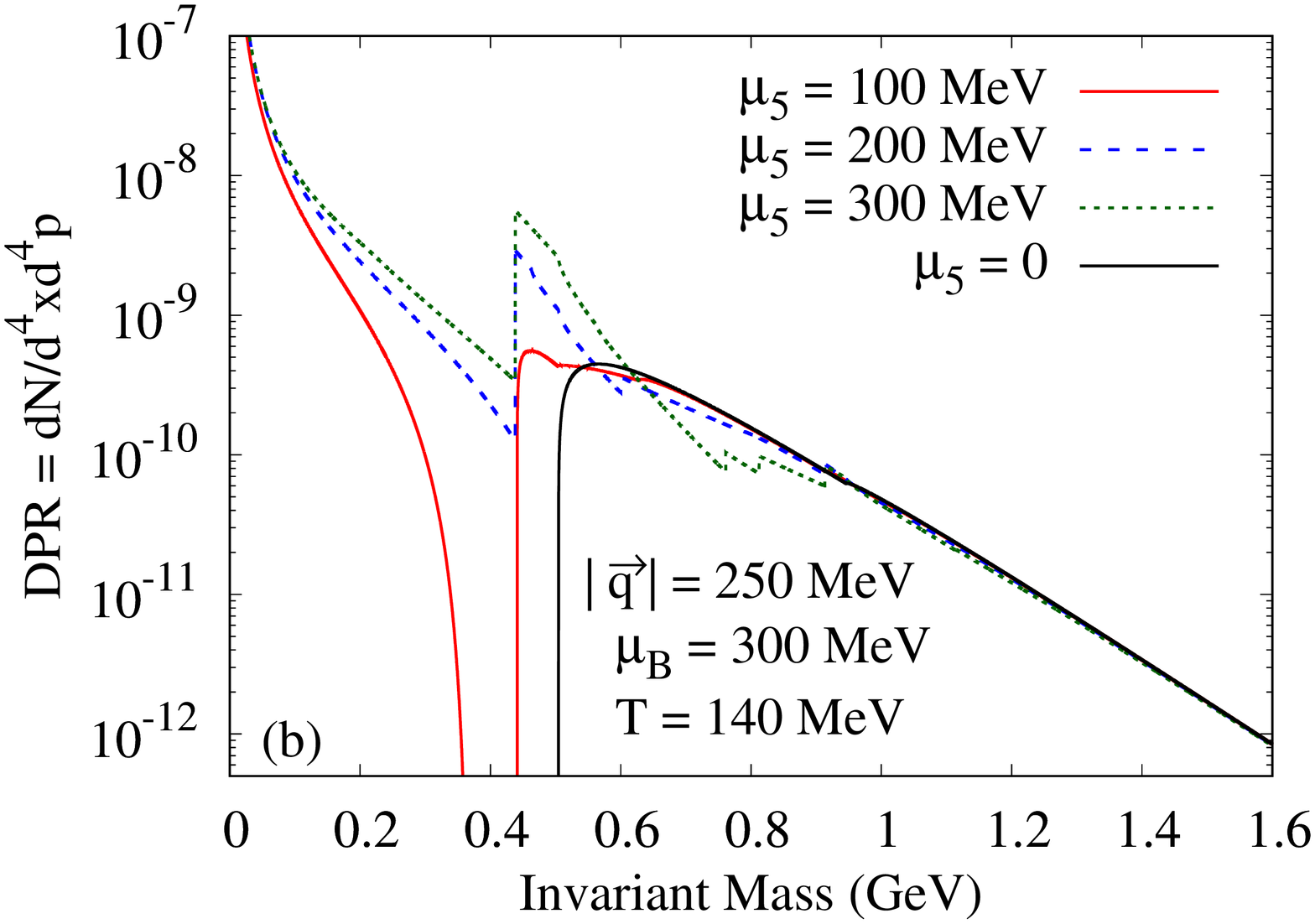}
		\includegraphics[angle = -90, scale = 0.233]{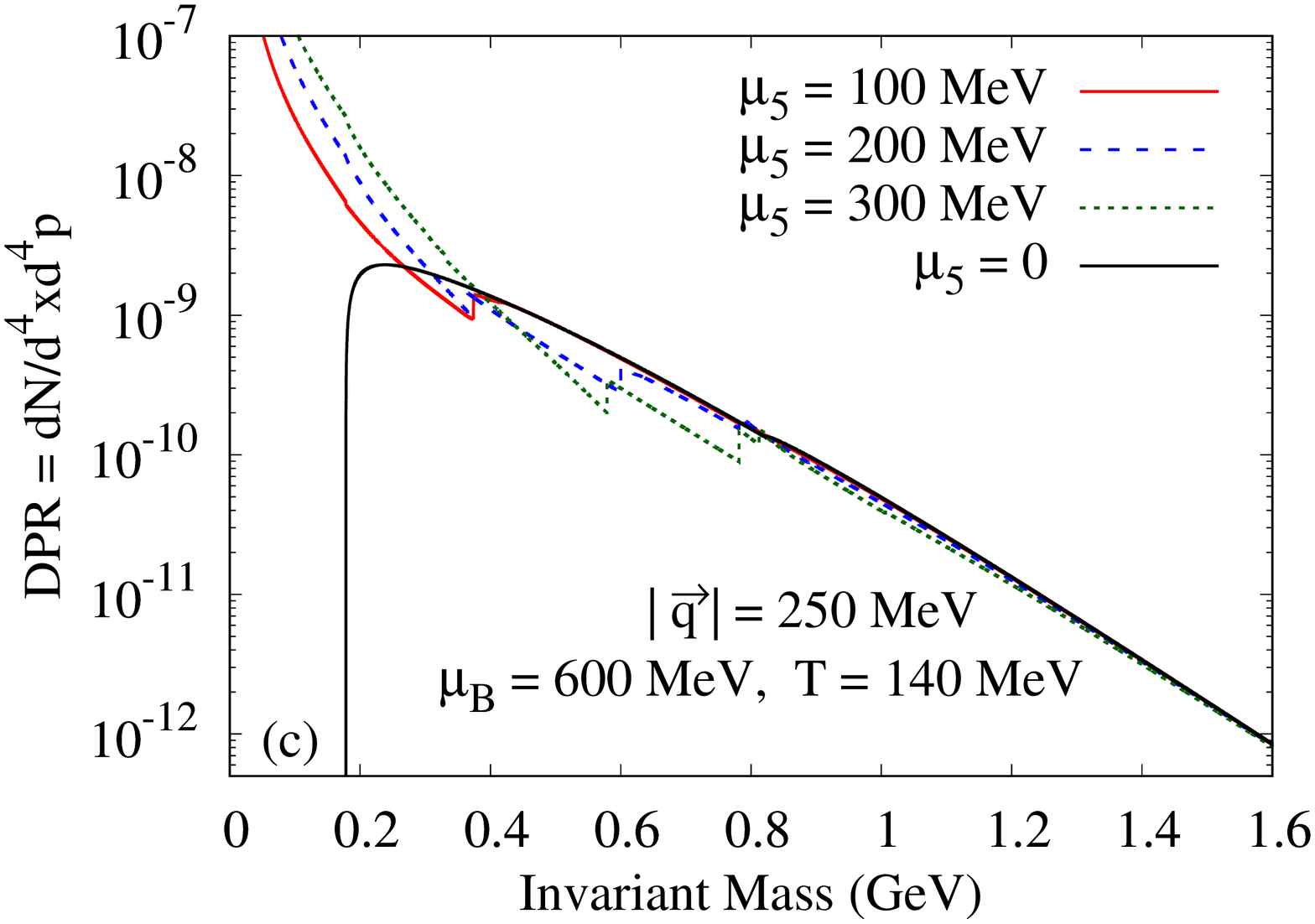}
		\caption{(Color Online)  Dilepton production rate at $ \qvec = 250$ MeV, $ T = 140 $  at (a) $ \mu_B  = 0 $ (b) $ \mu_B  = 300 $ MeV and (c) $ \mu_B  = 600 $ MeV respectively for different values of $ \mu_5 $. The DPR at $ \mu_5 = 0 $ are also shown for comparison. }
		\label{Fig_DPR1}		
	\end{figure}

Next, we present the numerical results for DPR from a hot and dense medium in presence of chiral imbalance.  Note that, all the results shown hereafter are obtained ignoring the lepton mass i.e. $ m_L = 0 $ and considering the momentum $ \qvec = 250  $~MeV. In Fig.~\ref{Fig_DPR1}(a) we have depicted DPRs as a function of invariant mass at $ T = 140 $~MeV and $ \mu_B = 0 $  for different values of CCP. The DPR at $ \mu_5 = 0 $ (black solid line) is also shown for comparison. For $ \mu_5  = 0$ case, DPR is restricted to low values of invariant mass as already discussed in Sec.~\ref{subsection_mu5_0}. The non-trivial contributions starts from just below $ 600 $ MeV owing to the Unitary-I cut threshold $ \sqrt{q} \ge 2M_u $. Now from Fig.~\ref{Fig_Mass} one can observe that, constituent mass of u-quark at $ T= 140 $ MeV is $ \sim 290 $ MeV. This explains the behaviour in case of vanishing CCP. However, in case of $ \mu_5  = 100 $ MeV a significant enhancement in the DPR can be seen owing to non-trivial Landau cut contributions.  From Eq.~\eqref{Table_Kinematic_Region}, one can observe that, the Landau cut contribution will be finite for physical dileptons when $ q_0 \le \qvec +2 \mu_5 $. This can be rewritten as $ \sqrt{q^2} \le 2\sqrt{\mu_5^2 + \qvec \mu_5}$. For $ \mu_5  = 100$ MeV, $ 2\sqrt{\mu_5^2 + \qvec \mu_5} \simeq 0.374 $ GeV. So around that value of invariant mass the Landau cut contribution should end, which can be seen in Fig.~\ref{Fig_DPR1}(a) (solid red line). Moreover, the unitary cut contributions starts at a lower value of invariant mass than $ \mu_5  = 0$ case although the constituent mass $ M_u $ is higher due to presence of finte CCP. To understand this let us concentrate on the Unitary cut thresholds for production of physical dileptons. From Eq.~\eqref{Table_Kinematic_Region}, this is expressed as $ q^0 \ge \sqrt{(\qvec-2\mu_5)^2+4M_f^2} $, which can be simplified further to arrive at $ \sqrt{q^2} \ge 2 \sqrt{ \mu_5^2 + M_f^2 - \mu_5 \qvec } $.  At $ T = 140 $ MeV, the constituent mass of the u-quark is $ \simeq  283$ MeV. So for $ \mu_5 = 100$ MeV, the Unitary cut contribution starts at $ \sqrt{q^2} \simeq 0.51 $ GeV, which is evident from Fig.~\ref{Fig_DPR1}(a).  This indicates that a chirally asymmetric medium can induce pair production at comparatively lower values of invariant mass as discussed in Sec.~\ref{subsection_mu5_ne_0}. It also explains the forbidden gap between Landau and Unitary cut where dilepton production ceases to occur. Now, from the above two discussions about the kinematic domain of Landau (Unitary) cut, it is clear that as we increase $ \mu_5 $ the threshold for dilepton production move towards higher (lower) values of invariant mass. Consequently, for higher values of CCP i.e. $ \mu_5  = 200 $ and $ 300 $ MeV, the Unitary and Landau cut contributions merge with each other resulting a continuous spectrum of dileptons for the whole range of invariant mass. In Fig.~\ref{Fig_DPR1}(b)  we have presented DPRs as a function of invariant mass at $ T = 140 $~MeV considering a finite baryon density ($ \mu_B = 300 $ MeV) for four different values of $ \mu_5 $. Notice that, for $ \mu_5 =100 $ MeV plot (solid red line) the Landau cut threshold ends at the same value as in zero baryon density case. This is understandable from the fact that the Landau cut threshold only depends on $ \mu_5 $.  However, the Unitary cut threshold, which is directly related to the constituent mass of quarks move towards lower values of invariant mass. This can be explained from Fig.~\ref{Fig_Mass} where one can observe that for finite values of $ \mu_B $ the (pseudo) chiral transition temperature moves towards the lower values of $ T $,  resulting in a decrease in the magnitude of $ M_u $. In Fig.~\ref{Fig_DPR1}(b), we find a continuous dilepton spectrum for higher values of $ \mu_5 $ which can be understood in a similar fashion as discussed earlier. In Fig.~\ref{Fig_DPR1}(c), we have considered an even higher baryon density ($ \mu_B  = 600 $ MeV) keeping all the other parameters same as Figs.~\ref{Fig_DPR1}(a) and (b). From the plot of vanishing CCP case, it is evident that, constituent mass of low lying quarks have decreased further such that, the Unitary cut  threshold for $ \mu_5 = 0 $ is already below the Landau cut threshold for $ \mu_5 = 100 $ MeV. As a consequence, we get a continuous spectrum of dileptons for the whole range of invariant mass for all values of CCP shown in the figure.
  
 \begin{figure}[h]
	\includegraphics[angle = -90, scale = 0.233]{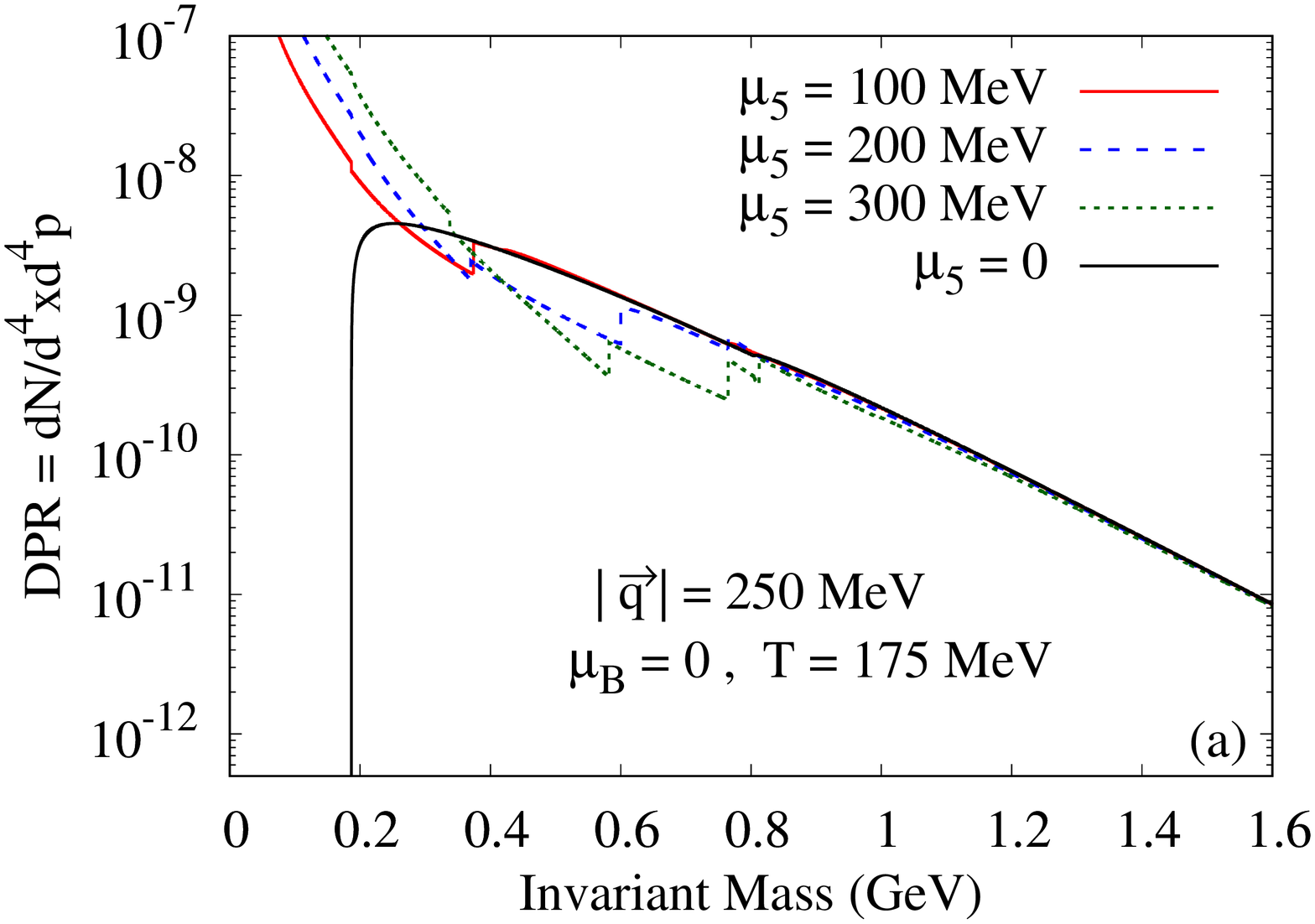}
	\includegraphics[angle = -90, scale = 0.233]{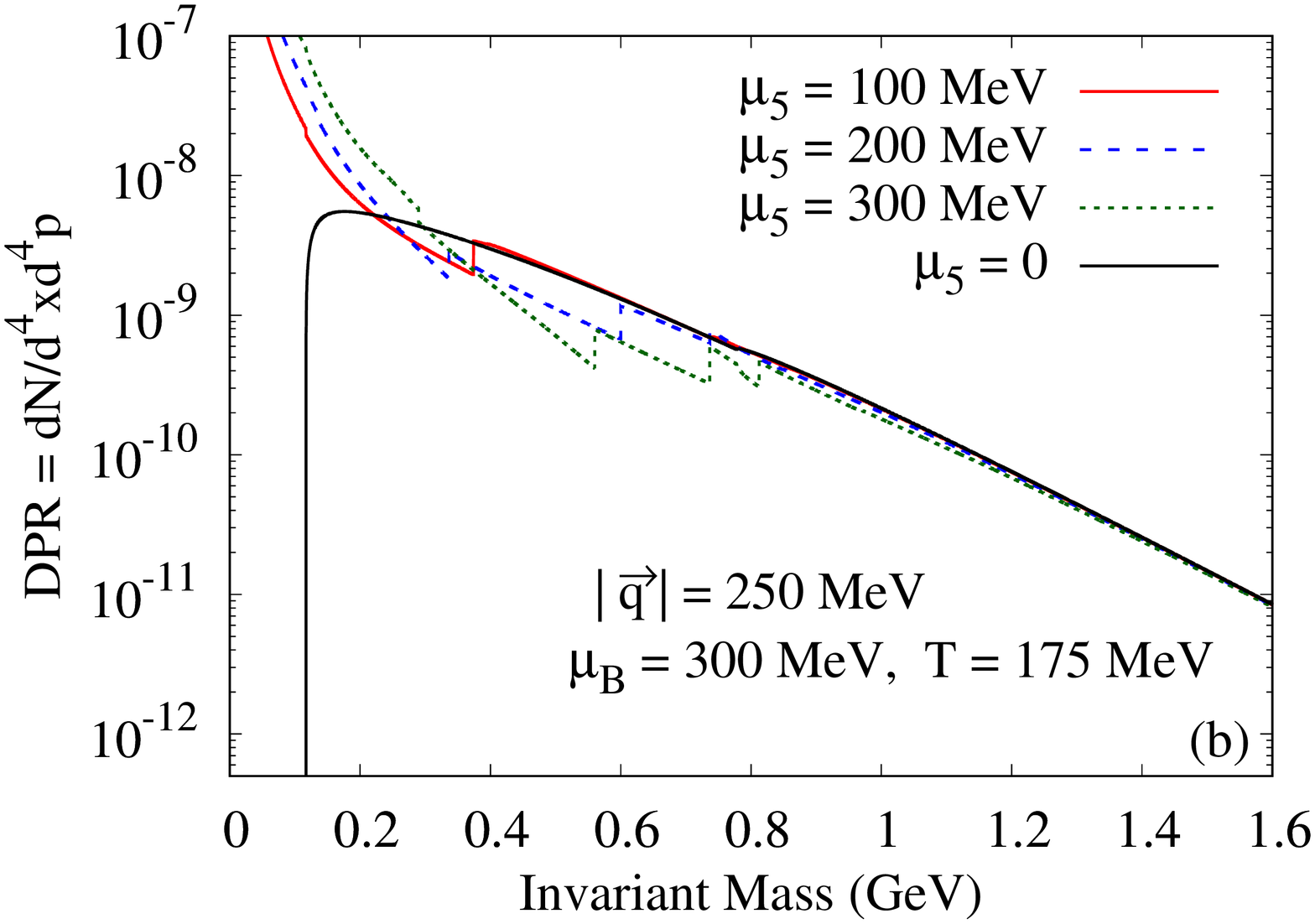}
	\includegraphics[angle = -90, scale = 0.233]{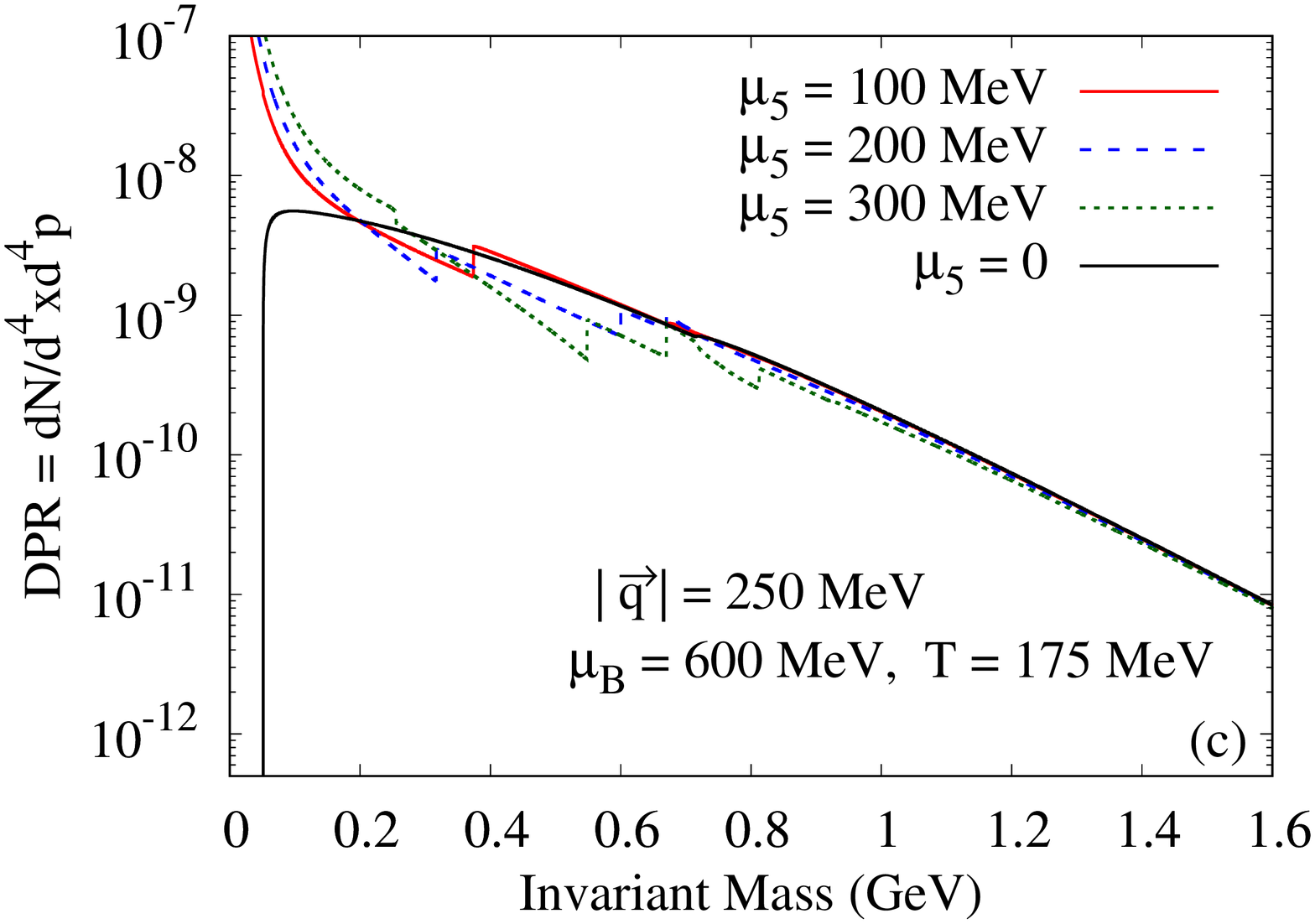}
	\caption{(Color Online)  Dilepton production rate at $ \qvec = 250$ MeV, $ T = 175 $  at (a) $ \mu_B  = 0 $ (b) $ \mu_B  = 300 $ MeV and (c) $ \mu_B  = 600 $ MeV respectively for different values of $ \mu_5 $. The DPR at $ \mu_5 = 0 $ are also shown for comparison. }
	\label{Fig_DPR2}		
\end{figure}
\begin{figure}[h]
	\includegraphics[angle = -90, scale = 0.233]{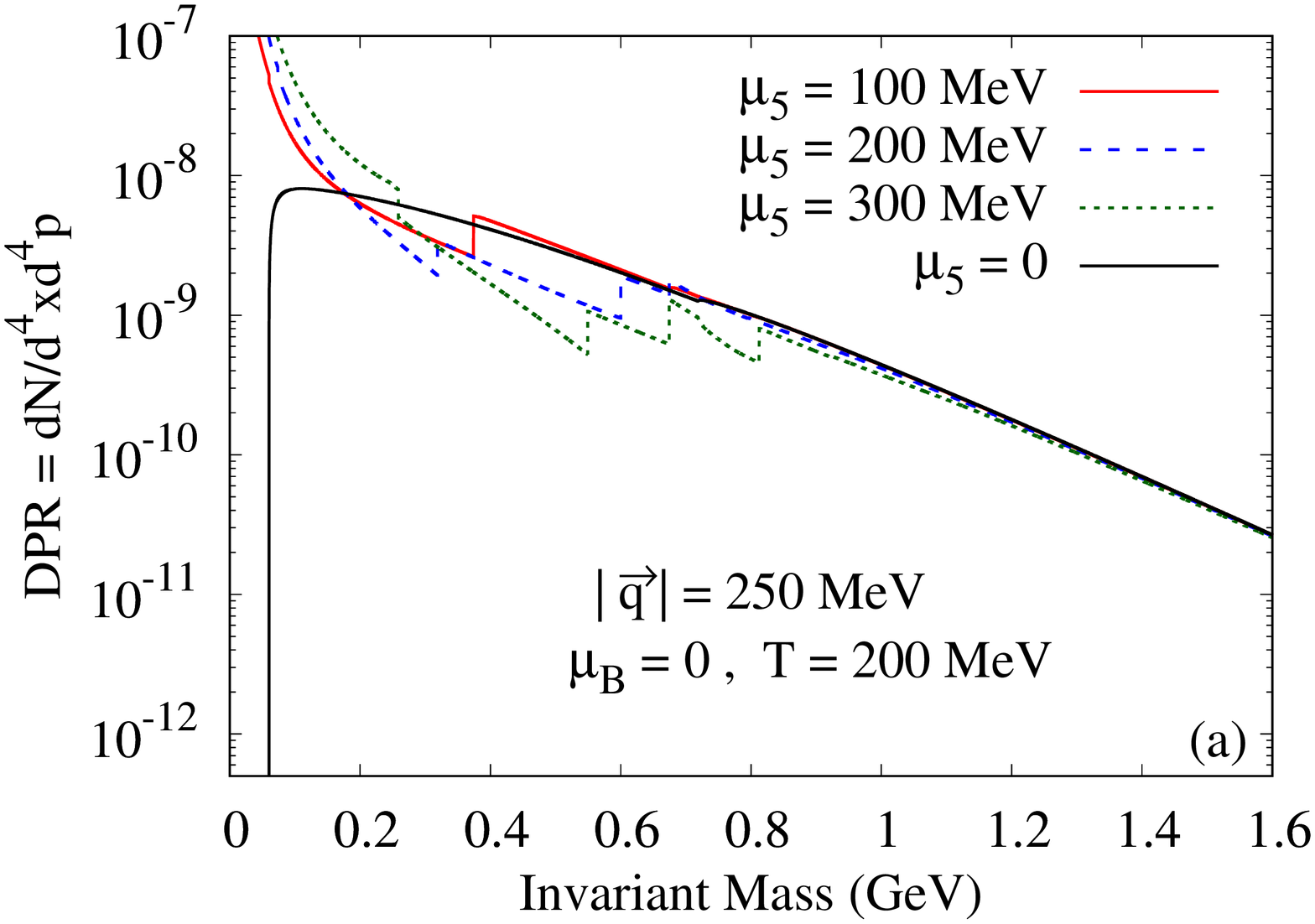}
	\includegraphics[angle = -90, scale = 0.233]{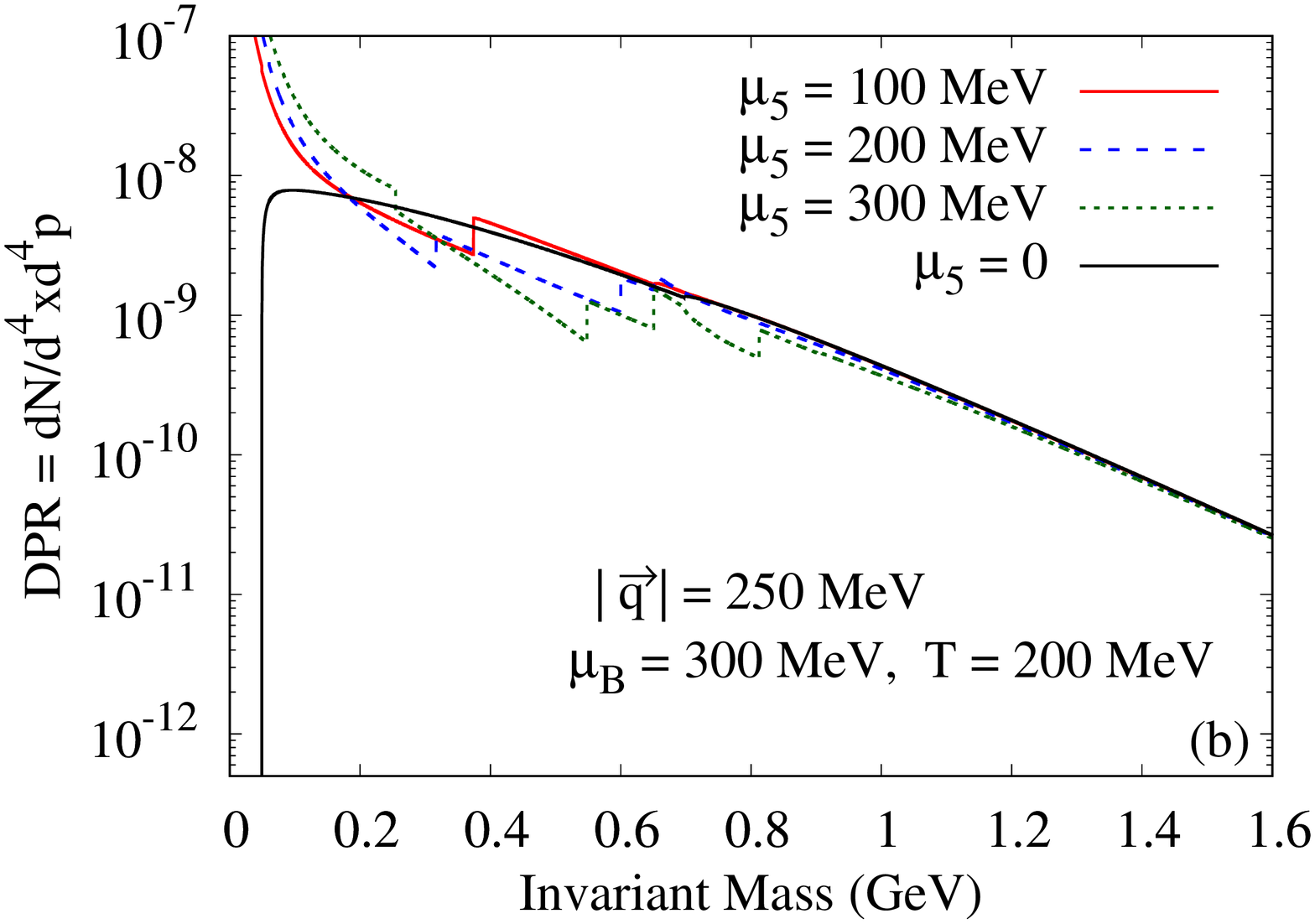}
	\includegraphics[angle = -90, scale = 0.233]{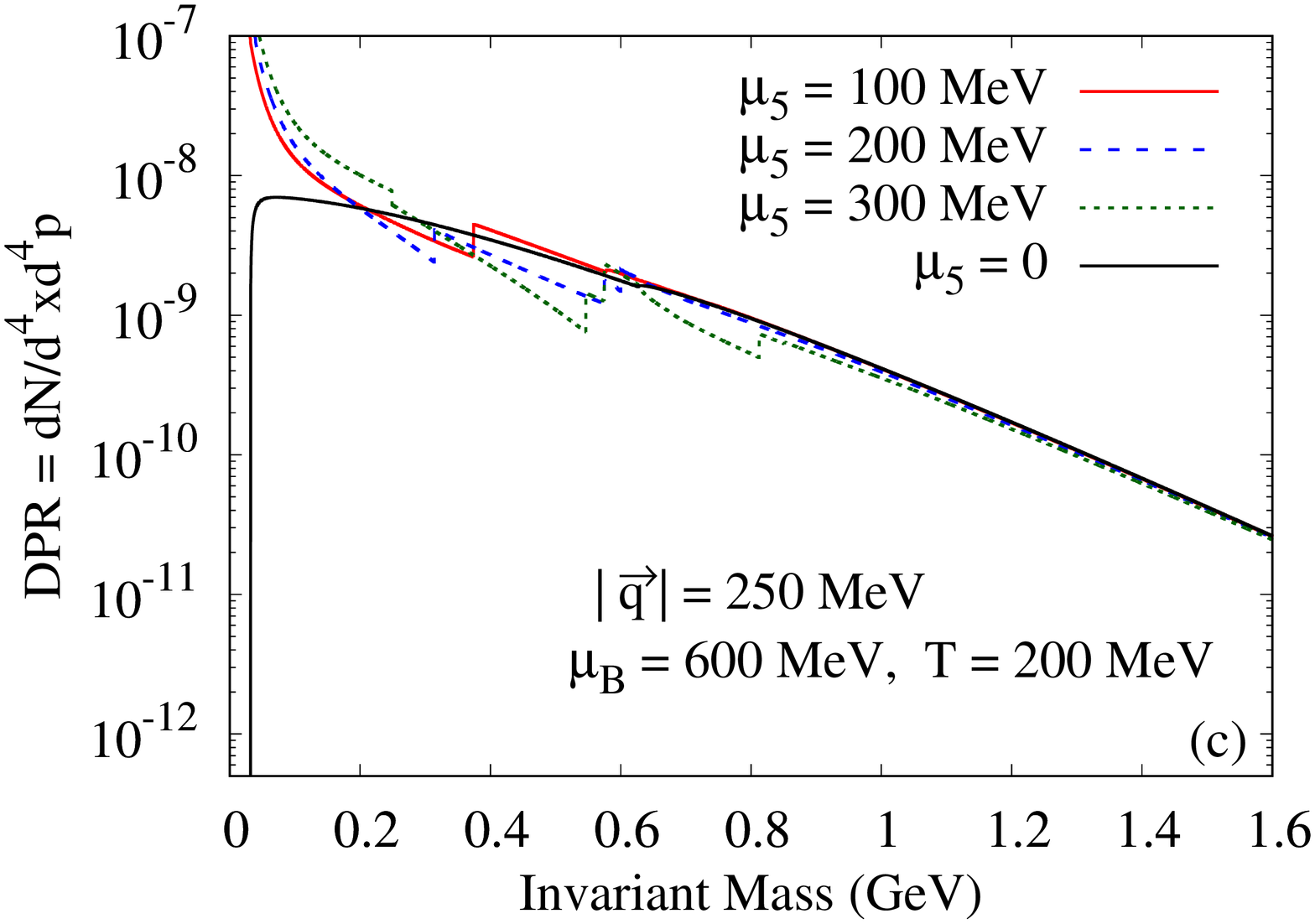}
	\caption{ (Color Online) Dilepton production rate at $ \qvec = 250$ MeV, $ T = 200 $  at (a) $ \mu_B  = 0 $ (b) $ \mu_B  = 300 $ MeV and (c) $ \mu_B  = 600 $ MeV respectively for different values of $ \mu_5 $. The DPR at $ \mu_5 = 0 $ are also shown for comparison. }
	\label{Fig_DPR3}		
\end{figure}

In Figs.~\ref{Fig_DPR2}(a), (b) and (c) we have depicted DPRs as a function of invariant mass at $ T = 175 $~MeV for different values of CCP for $ \mu_B  = 0, ~300 $ and 600 MeV respectively. Considering the zero baryon density case (Fig.~\ref{Fig_DPR2}(a)), it can be seen that, the Unitary cut  threshold for $ \mu_5 = 0 $ is already below the Landau cut threshold for $ \mu_5 = 100 $ MeV (note that, since the Landau cut contribution only depends of $ \mu_5 $, it will be same as the previous case). Moreover, from Fig.~\ref{Fig_Mass}, it is clear that, at $ T = 175$ MeV, the constituent mass of low lying quarks are always below $ \lesssim 150 $ MeV. As a result, at $ T=175 $ MeV, for all finite $ \mu_5 $ case shown in the figure, the Landau and Unitary cut contribution will merge and a continuous spectrum of dilepton will emerge as evident from Figs.~\ref{Fig_DPR2} (a), (b) and (c).

A similar trend is visible in  Figs.~\ref{Fig_DPR3}(a), (b) and (c) where the DPR is plotted for $T=200$ MeV. Here, because of restoration of chiral symmetry the constituent mass of the low lying quarks goes to the bare mass limit (see Fig.~\ref{Fig_Mass}). Consequently, the threshold of unitary cut starts at smaller values of the invariant mass. Moreover, since the temperature considered in this case is higher than all the previous cases the increase in thermal phase space results in a significant enhancement in the overall magnitude of DPRs  when compared to Figs.~\ref{Fig_DPR1}(a), (b) and (c))~\cite{Ghosh:2018xhh,Ghosh:2020xwp,Chaudhuri:2021skc}.

We end this section with a discussion on the experimental observation of the effects of chiral imbalance on the dilepton spectra. Keeping in mind that dileptons are emitted at all stages of the collision, the DPR from quark matter as well as those from $\rho$ and $\omega$ decays and other hadronic reactions  have to be evolved in space-time using (1+3)d hydrodynamics (with  $\mu_5\ne 0$). As discussed earlier, chiral imbalance may be created in HICs locally which leads to a continuous spectrum with a distinct shape in the low invariant mass region of the DPR from quark matter. It should be noted that the creation of a local domain with chiral imbalance changes from event to event and the event average value of any $\mu_5$-dependent observable is zero. Thus, the analysis must be done on an event by event basis. Now let us consider an event in which a domain with chirally imbalanced quark matter is created. Due to the hydrodynamic expansion it will cool and undergo a transition to hadronic matter which presumably will also be in $P$ and $CP$ odd phase. If that be the case, there will be effect of chiral imbalance on the DPR coming from resonance decay and hadronic reactions as well. Andrianov et. al. \cite{Andrianov:2012hq} have predicted that such matter will produce an excess of dileptons in the $\rho-\omega$ resonance region. Together with the enhancement of DPR from quark matter in the low invariant mass region  (i.e. non-zero yield in the invariant mass range $M \sim 0.2 - 0.4$ GeV with $\mu_5 \sim 200$ MeV or higher) seen in this work, the total dilepton yield could explain the observed low mass enhancement seen in the PHENIX experiment~\cite{Liao:2014ava} for central collisions which cannot be explained by invoking the temperature and density dependent medium modifications of hadronic spectral functions alone.

%~~~~~~~~~~~~~~~~~~~~~~~~~~~~~~~~~~~~~~~~~~~~~~~~~~~~~~~~~~~~~~~~~~~~~~~~~~~~~~~~~~~~~~~~~~~~~~~~~~~~~~~~~	
	\section{SUMMARY \& CONCLUSION}
	In this work we studied the dilepton production rate from hot and dense chirally asymmeteric matter likely to be produced in relativistic HICs. The electromagnetic spectral function which is the principal component in the DPR is found to be modified due to the presence of a CCP. The constituent quark mass which appears in the in-medium propagator is evaluated in a self-consistent manner using a 3-flavour NJL model and is also non-trivially affected by the CCP. Specifically, in presence of $ \mu_5 $ the quark condensate is found to be enhanced at small values of $ T $ while it is hindered at high values of temperature. We have also analyzed the complete analytic structure of the spectral	function of the chirality imbalanced medium in the complex energy plane and have found a nontrivial Landau cut in the physical kinematic region signifying additional scattering processes in the medium; a purely finite CCP effect. As a consequence the DPR	acquires contributions from both the unitary and Landau cuts for production of dileptons with positive energy and time-like four momentum. Owing to the emergence of the Landau cut the DPR is highly enhanced in the low invariant mass region compared to the case with a vanishing CCP. It is also found that the Landau cut threshold is independent of the constituent quarks mass though the unitary cut threshold has a non-trivial dependence on both $ M_f $ and $ \mu_5 $. For small values of $ T $ and $ \mu_5 $ a forbidden gap exists between Unitary and Landau cuts where production of dileptons is kinematically restricted. However, as we increase the CCP at fixed values of $ T $ and $ \mu_B $ the Landau cut threshold moves towards higher invariant mass. As a result the forbidden gap  keeps shrinking and eventually merges with each other producing a continuous spectrum of dileptons as a function of invariant mass even for small values of temperature where chiral symmetry is still broken. For higher values of temperature and/or BCP the forbidden gap completely disappears and a significant rise in dilepton production rate is observed owing to the availability of larger thermal phase space. 

%~~~~~~~~~~~~~~~~~~~~~~~~~~~~~~~~~~~~~~~~~~~~~~~~~~~~~~~~~~~~~~~~~~~~~~~~~~~~~
\section*{Acknowledgments}
NC, SS and PR are funded by the Department of Atomic Energy (DAE), Government of India. SG is funded by the Department of Higher Education, Government of West Bengal, India.

	\bibliography{DL_mu5}

\end{document}